%% file: paper_methods.tex
\documentclass[a4paper,12pt]{article}
\usepackage[paper=a4paper,left=25mm,right=25mm,top=25mm,bottom=25mm]{geometry}

\usepackage[english]{babel}
\usepackage[utf8]{inputenc}
\usepackage{longtable}
\usepackage{comment}
\usepackage{graphicx}
\usepackage{natbib}
\usepackage{xcolor}
\usepackage{setspace}
\usepackage{soul}

\usepackage{pdflscape}

\usepackage{chngcntr}

\usepackage{booktabs} 
\usepackage{colortbl} 
\usepackage{array} 

\usepackage{array}
\usepackage{arydshln} 
\usepackage{amsmath}

\usepackage[bottom]{footmisc}
\usepackage{scalerel}
\usepackage{tikz}

\usepackage{amssymb}

\usepackage{authblk}

\usepackage{orcidlink}
\usepackage{url}

\definecolor{JournalBlue}{RGB}{0, 12, 146}

\usepackage[colorlinks=true, linkcolor=JournalBlue, filecolor=JournalBlue, urlcolor=JournalBlue, pdfborder={0 0 0},citecolor=JournalBlue]{hyperref}

\title{A Supervised Machine Learning Approach for Assessing Grant Peer Review Reports
}

\date{\normalsize Version: \today}

\author[1]{\large Gabriel Okasa\,\orcidlink{0000-0002-3573-7227}}
\author[2]{Alberto de León\,\orcidlink{0009-0002-0401-2618}}
\author[1]{Michaela Strinzel\,\orcidlink{0000-0003-3181-0623}} 
\author[1]{Anne Jorstad\,\orcidlink{0000-0002-6438-1979}}
\author[1]{Katrin Milzow\,\orcidlink{0009-0002-8959-2534}}
\author[1,3,4]{Matthias Egger\,\orcidlink{0000-0001-7462-5132}}
\author[2]{Stefan Müller\,\orcidlink{0000-0002-6315-4125}\thanks{Corresponding author:  
Stefan Müller (\href{mailto:stefan.mueller@ucd.ie}{stefan.mueller@ucd.ie}). \medskip \\ 
Previous versions of this paper were presented at workshops of the Swiss National Science Foundation in Bern, the COMPTEXT conference in Amsterdam, the SPIRe Seminar Series at University College Dublin, and the REvaluation conference in Vienna. We would like to thank the workshop and conference participants for their valuable feedback, and Sarah King, Lorcan McLaren, and Jihed Ncib for outstanding research assistance. All errors are our own.
}}

\affil[1]{\normalsize Swiss National Science Foundation}
\affil[2]{University College Dublin}
\affil[3]{University of Bern}
\affil[4]{University of Bristol}

\begin{document}

\definecolor{snsfblue}{HTML}{3D7D9F}
\definecolor{snsfgreen}{HTML}{71B294}
\definecolor{snsfred}{HTML}{D9534F}

\newgeometry{left=3.4cm, right=3.4cm, top=15mm,bottom=25mm}
\maketitle

\vspace{-0.5cm}

\begin{abstract}
\noindent
Peer review in grant evaluation informs funding decisions, but the contents of peer review reports are rarely analyzed. In this work, we develop a thoroughly tested pipeline to analyze the texts of grant peer review reports using methods from applied Natural Language Processing (NLP) and machine learning. We start by developing twelve categories reflecting content of grant peer review reports that are of interest to research funders. This is followed by multiple human annotators' iterative annotation of these categories in a novel text corpus of grant peer review reports submitted to the Swiss National Science Foundation. After validating the human annotation, we use the annotated texts to fine-tune pre-trained transformer models to classify these categories at scale, while conducting several robustness and validation checks. Our results show that many categories can be reliably identified by human annotators and machine learning approaches. However, the choice of text classification approach considerably influences the classification performance. We also find a high correspondence between out-of-sample classification performance and human annotators’ perceived difficulty in identifying categories. Our results and publicly available fine-tuned transformer models will allow researchers and research funders and anybody interested in peer review to examine and report on the contents of these reports in a structured manner. Ultimately, we hope our approach can contribute to ensuring the quality and trustworthiness of grant peer review.

\end{abstract}

\medskip

\thispagestyle{empty}

\restoregeometry

\pagebreak

\doublespacing

\setcounter{page}{1}

 
\section{Introduction}

Research funders rely on external peer review to evaluate proposals for scientific merit, methodological rigor, feasibility, and potential impact. Despite the central role of peer review in grant evaluation and funding decisions, analyses of peer review reports are rare and challenging. Science has become more open, with journals increasingly publishing peer review reports and authors' responses alongside the articles. In contrast, grant peer review reports are confidential to allow reviewers to comment freely and to protect ideas presented in grant applications. Interestingly, a recent study suggested that "researchers trust grant reviewers far less than they trust journal peer reviewers..." \citep[28]{langfeldt24}, likely due to a perception that the grant evaluation process lacks transparency.

Analyzing the content of grant peer review reports could provide several benefits, including an objective assessment of review quality to gauge adherence to guidelines, ensuring thoroughness, fairness, and consistency in panel discussions, and improving decision-making. Such analyses may enhance the transparency of evaluation processes and identify areas for improvement in reviewer training. Prior work has investigated potential biases in peer review based on the gender and racial disparities of authors and reviewers \citep{erosheva2020nih,squazzoni2021peer}. Various methodologies have emerged in recent years for analyzing peer-review reports, focusing on content, sentiment, and argument structures. Early studies primarily used linguistic and sentiment analysis to explore the relationship between reviewer language and review scores, offering foundational insights \citep{luo22}. While straightforward, these methods struggle to capture more nuanced information. More advanced approaches combine machine learning with qualitative techniques, such as clustering analysis to assess alignment with evaluation criteria \citep{hren2022}. Recently, machine learning models like random forests and transformers have been applied for more detailed classification of peer review texts \citep{williams23, severin23, kuznetsov2024nlp}. A key development is the increasing use of argument detection models, which leverage deep learning to classify and segment argumentative propositions within peer reviews \citep{Fromm2021, Lawrence2020}. The latest innovations focus on Large Language Models (LLMs), such as GPT-4 or Llama, for generating labels and automating quality assessments, significantly improving efficiency and accuracy \citep{Pelaez2024}. 

In this study, we integrate human annotation with transformer-based machine learning to create a tool for evaluating the characteristics of grant peer review reports submitted to the Swiss National Science Foundation (SNSF) between 2016 and 2024. We present each step in detail, report initial findings, and discuss the potential of machine learning as a tool for gaining insights into the grant peer review process. The code scripts for the conducted analyses are publicly available on GitHub\footnote{\url{https://github.com/snsf-data/ml-peer-review-analysis}.} and the fine-tuned models are shared on the Hugging Face Hub\footnote{\url{https://huggingface.co/snsf-data}.} to enable others to benefit from our work.

\section{Methods}

\begin{figure}[h!]
\caption{Pipeline of identifying, classifying, and validating textual characteristics in grant peer review reports} \label{fig:diagram}
\centering
\includegraphics[width=0.8\textwidth]{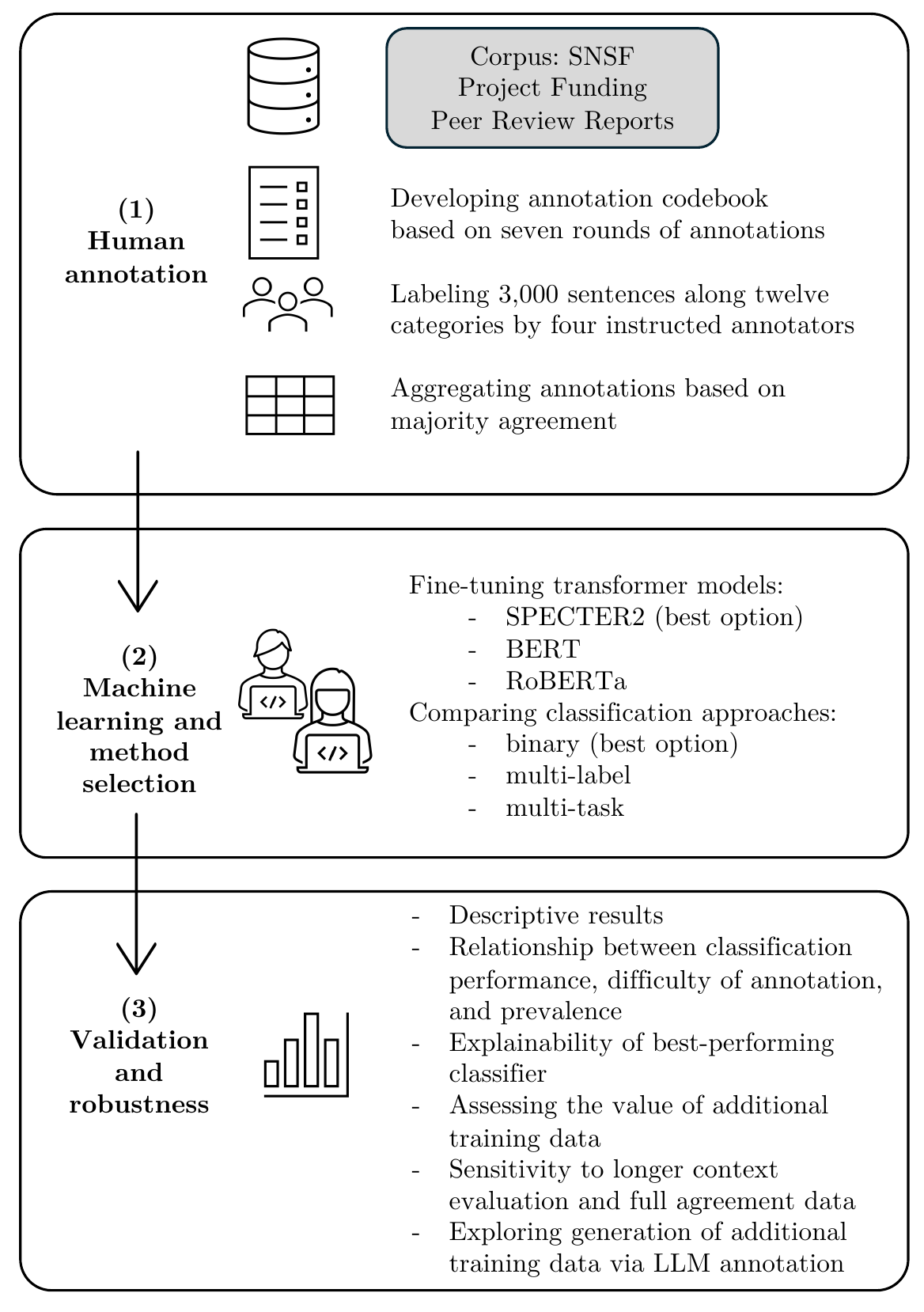}
\end{figure}

The approach involves three key steps: (1) human annotation; (2) machine learning and method selection; (3) and validation and robustness. The steps are summarized in Figure \ref{fig:diagram} and outlined below.
The training of the machine learning models was performed locally without access to the internet to prevent any potential data leakage or network interference.

\subsection{Setting}

The study relies on a corpus of 47,522 external peer review reports written in English for 13,653 applications submitted to the SNSF's Project Funding scheme between October 2016 and April 2024. Project Funding is the largest SNSF funding scheme, open to research proposals from all academic disciplines. In 2023, the SNSF funded 2,928 grants with CHF 1.2 billion, of which 1,009 grants (CHF 580 million) were in the Project Funding scheme.\footnote{See the SNSF's DataPortal for more details: \url{https://data.snf.ch/key-figures/funding-instruments?s5=0&s2=1}} The SNSF evaluation procedure consists of several evaluation steps. The peer review reports analyzed in this study constitute the ``External review,'' which is the first part of the SNSF's scientific evaluation process (see Figure \ref{fig:evaluation_proc_main} in Appendix for details). At least two external experts review research proposals against a set of  evaluation criteria specified by the SNSF: (1) \textit{Scientific relevance, originality and topicality}, (2) \textit{Suitability of methods and feasibility}, (3) \textit{Applicant’s scientific track record and expertise}. Proposals tagged as `use-inspired' by the applicants are assessed by taking into account their broader impact as an evaluation sub-criterion: (1a) \textit{Broader impact}. External reviewers provide their assessment in text boxes for each criterion and an overall comment, which all address the proposal's specific strengths and weaknesses, resulting in up to 5 distinct text comments in total. (see Figure \ref{fig:evaluation_proc_main} in Appendix for details on the evaluation criteria). 

 Alongside the structured written report, external reviewers rate the proposals on a numeric scale. The SNSF evaluation panels then discuss and rate the proposals taking into account the external peer review reports. Funding decisions are based on the panel's ranking of proposals \citep{heyard2022}.  

We cleaned and anonymized the raw texts of the review reports. This included removing redundant characters such as HTML code and the names of applicants, co-applicants and project partners. The data management plan underlying this work provides a detailed description of data protection.\footnote{ \url{https://doi.org/10.46446/DMP-peer-review-assessment-ML}} We limited the analysis to reviews submitted in English. Thus, our analysis does not include 10.2\% of the originally submitted reviews conducted in different languages (mostly German and French), all of which come from the humanities and social sciences.

\subsection{Codebook Development and Annotation}

We conducted eleven distinct annotation rounds, during which we developed and refined a comprehensive annotation codebook with instructions.\footnote{ \url{https://doi.org/10.46446/Codebook-peer-review-assessment-ML}} The final version of the codebook includes twelve distinct categories, each relevant to the criteria specified by the SNSF. These categories are applied at the sentence level of each review and are grouped into three dimensions: 1)\textit{ Evaluation Criteria}, 2) \textit{Focus}, and 3) \textit{Statement Type and Reasoning}.

The dimension \textit{Evaluation Criteria} examines to what extent the reviewer applied the criteria outlined above. \textit{Focus} captures the aspect the reviewer is focusing on: the applicant (quantitative and general comments), the proposal (such as the project plan), or the methodological framework. \textit{Statement Type and Reasoning} evaluates whether the statement is positive or negative, whether it is substantiated with evidence, and whether the reviewer offers suggestions for improvement. Table \ref{tab:overview_categories} provides a detailed overview of the categories.

\begin{table}[h!]
\centering
\caption{Overview of the 12 categories used in the final annotation rounds}\label{tab:overview_categories}
\begin{tabular}{|p{7cm}|p{7cm}|}
\hline
\multicolumn{2}{|c|}{\cellcolor{snsfblue}\textbf{\textcolor{white}{Evaluation Criteria}}} \\
\multicolumn{2}{|c|}{\cellcolor{snsfblue}\textit{\textcolor{white}{Which criteria does the reviewer apply?}}} \\\hline
\textit{Track record and expertise} & \textit{Relevance, originality, and topicality} \\ 
Does the sentence address the scientific qualifications of the applicant(s)/team? & Does the sentence address the scientific relevance/originality/topicality of the proposed research project? \\ \hline
\textit{Suitability} & \textit{Feasibility}  \\
Does the sentence address the suitability of the methods to be used within the proposed research project?  & Does the sentence address the feasibility of the proposed research project? \\ \hline
\multicolumn{2}{|c|}{\cellcolor{snsfgreen}\textbf{\textcolor{white}{Focus}}}\\
\multicolumn{2}{|c|}{\cellcolor{snsfgreen}\textit{\textcolor{white}{Which part(s) of the application documents does the reviewer focus on?}}}\\ \hline
\textit{Applicant: Quantity} & \textit{Applicant} \\ 
Does the sentence use quantitative indicators to describe the applicant(s) or team?  & Does the sentence address the applicant(s)/team or their qualifications, without mentioning quantitative indicators? \\ \hline
\textit{Proposal} & \textit{Method} \\ 
Does the sentence address the proposal or specific parts of it, as opposed to the applicant(s) or context beyond the proposal (such as the research field or the funding scheme’s objectives etc.)?  & Does the sentence address the methods to be used in the proposed research project? \\ \hline

\multicolumn{2}{|c|}{\cellcolor{snsfred}\textbf{\textcolor{white}{Statement Type and Reasoning}}}\\
\multicolumn{2}{|c|}{\cellcolor{snsfred}\textit{\textcolor{white}{Does the reviewer justify their evaluations and/or offer suggestions?}}} \\  \hline
\textit{Positive statement} & \textit{Negative statement} \\ Is the sentence itself a positive statement or does it contain a positive statement?   & Is the sentence itself a negative statement or does it contain a negative statement? \\ \hline
\textit{Suggestion} & \textit{Rationale} \\ 
Does the sentence suggest how to improve the proposal?  & Does the sentence provide rational supporting the positive or negative statement? \\ \hline
\end{tabular}
\end{table}

Each annotation round relied on randomly sampled sentences from the entire text corpus. Table \ref{tab:appendix_overviewrounds} in the Appendix provides a detailed overview of each annotation round.  The sampled sentences are representative of the three main research domains.\footnote{See the SNSF discipline list: \url{https://www.snf.ch/SiteCollectionDocuments/allg_disziplinenliste.pdf}.} 38.3\% of randomly sampled review sentences are from proposals submitted to the research domain of Mathematics, Information Technology, Natural- and Engineering Sciences (MINT). A similar share of review sentences (36.9\%) comes from a review submitted to the Life Sciences (LS). The remaining 24.7\% of review sentences are based on proposals in the Social Sciences and Humanities (SSH). These percentages correspond to the distribution of reviews across disciplines in the full-text corpus.

The human annotators classified each sentence across the twelve categories. The categories are not mutually exclusive and were annotated independently of each other. The only exception is \textit{Rationale}, which was only considered if the annotator first identified a positive or a negative statement.

The linguistic backgrounds of the annotators, two native English speakers and two non-native English speakers, reflect the context in which the data is created: while the language of the peer review reports is English, they are written and assessed by both native and non-native English speakers. The annotation codebook, with clear instructions and regular inter-coder reliability tests, ensured consistency and rigor throughout the process (see Table \ref{tab:appendix_overviewrounds}).

 The final annotation set comprised 3,000 sentences. The first 2,000 sentences were sampled randomly from the entire corpus of English reviews. Stratified sampling was used for the remaining 1,000 sentences. We randomly selected  200 sentences from each of the five text boxes reflecting the SNSF evaluation criteria (see Figure \ref{fig:evaluation_proc}) to obtain more sentences for under-represented categories, such as \textit{Applicant: Quantity}.

To validate annotation, we randomly assigned each sentence to three of the four annotators. Their annotations were aggregated based on majority agreement: if at least two of three annotators identified a particular characteristic in a sentence, it was classified as an example of that category. If only one or no annotator identified the characteristic, the sentence was classified as not belonging to that category. This consensus-based method reduces annotator bias \citep{benoit16}, improving the reliability of the annotated dataset.

Further, a comprehensive analysis of agreements and disagreements between annotators was conducted after each annotation round. This process facilitated a nuanced understanding of annotation discrepancies, enabling continuous refinement of the coding guidelines and improvement of annotation consistency. Appendix \ref{sec:examplesentences} lists five example sentences from each category with full agreement among annotators, while Appendix \ref{app:descriptives} provides an overview of the descriptive statistics for the sample of annotated sentences.

Next, we used the human-annotated sample of 3,000 sentences to train machine learning models that predict the category labels for the remaining review sentences in our dataset and for future incoming reviews. For this purpose, we relied on a pre-trained language model \citep{vaswani17, tunstall22} and fine-tuned it for the text classification task based on our annotated review sentences.

\subsection{Transformer Models}

We focus on the so-called BERT models (Bidirectional Encoder Representations from Transformers) developed at Google  \citep{devlin18}.  The BERT model is pre-trained using masked language modeling and next-sentence prediction based on a large text corpus. It learns via bidirectional representations, conditioning on both left and right context in the text sequence in all layers of the model \citep{devlin18}. Due to its strong performance and efficiency, BERT has become a foundational model for many NLP tasks, including text classification \citep{acheampong2021transformer}. The text corpus consists of the BookCorpus \citep{zhu15} and  English-language  Wikipedia. 
Variations of the original BERT model modify the pre-training by improving  the optimization or the training data. Given the scientific nature of our dataset, which consists of peer review reports, we follow the developments of SciBERT,  \citep{beltagy19}, a version of BERT pre-trained on scientific texts from SemanticScholar. SciBERT has been further enhanced by citation graphs, leading to the SPECTER model \citep{cohan20}. Specifically, we use SPECTER2 \citep{singh22}, an updated version of SPECTER, which outperforms the original BERT model on various NLP tasks in the scientific domain \citep[see e.g.][]{wolff2024enriched, Okasa2024}. We deploy the SPECTER2 model available on Hugging Face\footnote{\url{https://huggingface.co/allenai/specter2_base}.} via the \textsf{transformers} library in Python \citep{wolf20}. To test the effectiveness of SPECTER2, we also fine-tuned the original BERT model as a baseline, along with RoBERTa, a more robustly optimized version of BERT   \citep{Liuetal2019}.

\subsection{Fine-tuning}

The pre-trained SPECTER2 model can be used as-is to extract text representations via the CLS token  \citep{cohan20}. We not only extracted embeddings through the CLS token for input into a classifier, but also fine-tuned SPECTER2 on our classification task using the dataset of 3,000 annotated sentences.\footnote{We perform the fine-tuning using a GPU unit with 4 GB of memory (NVIDIA RTX A2000).}

Formally, consider $\{(X_i,y_i)\}_{i=1}^n$ to be an \textit{i.i.d.} random sample of size $n$ with $X_i$ being the $i$-th text sequence (sentence) and $y_i \in \{0,1\}$ being the corresponding outcome (category) with binary class labels. We are interested in predicting the conditional probability of the outcome given the input text sequence, i.e. $P[y_i=1 \mid X_i=x]$. We encode each input sequence $X_i$ into a vector representation via the text embedding function of SPECTER2, $\mu(\cdot)$, as follows
\vspace{-0.1cm}
$$h_i=\mu(X_i;\theta)$$
\vspace{-1.0cm}

\noindent where $h_i$ is the contextualized representation for the text sequence $X_i$ based on the CLS token and $\theta$ being the model parameters. We then feed the embeddings into a fully connected linear output layer with weights $W$ and bias $b$ to get the output scores, i.e. the logits, and apply a sigmoid activation function, $\sigma(\cdot)$, to obtain the class probabilities as
\vspace{-0.5cm}
$$\hat{y}_i=\sigma(W\cdot h_i + b)$$
\vspace{-1.35cm}

\noindent with $\hat{y}_{i} = \hat{P}[y_i=1 \mid X_i=x]$ being the predicted class probability for input $X_i$. The prediction of the class label for a text sequence $X_i$ is then obtained via thresholding as
\vspace{-0.25cm}
$$\hat{y}_i^{\mathcal{C}} = \mathbb{I}(\hat{y}_i \geq 0.5 )$$
\vspace{-1.25cm}

\noindent with $\mathbb{I}$ being an indicator function. In the course of the fine-tuning, the optimal values for parameters $\theta$, $w$ and $b$ are determined by optimizing the cross-entropy loss
\vspace{-0.3cm}
$$\mathcal{L}(\theta, w, b) = -\frac{1}{n}\sum_{i=1}^n \big( y_i \cdot \log(\hat{y}_i) + (1-y_i)\cdot \log(1-\hat{y}_i)\big).$$
\vspace{-1.0cm}

\noindent As such, the parameters of the linear output layer, $W$ and $b$,  are trained from scratch, whereas the parameters of the pre-trained SPECTER2 model, $\theta$, are fine-tuned.

Such fine-tuning requires selecting an optimizer and its hyperparameters to guide the learning process. We used the Adam optimizer with decoupled weight decay (AdamW), which improves generalization and offers computational advantages over stochastic gradient descent with momentum \citep{loshchilov17}. We follow recommendations from  \cite{devlin18}, using a learning rate of 2e-5, weight decay of 0.01, 3 epochs, and a batch size of 10. Table \ref{tab:settings} in the appendix provides a detailed overview of the fine-tuning hyperparameters.

\subsection{Text Classification Approach}

Several methods are available for classifying texts into various categories \citep[for an overview, see][]{minaee2021deep}. To identify the most effective approach, we empirically compared three approaches: binary, multi-task, and multi-label classification.

\subsubsection{Binary Classification}
The binary classification approach divides the task into 12 separate binary classification problems, with each model dedicated to a specific category. As a result, we fine-tune 12 separate models, one for a each outcome category. While this method is computationally intensive due to the need to fine-tune multiple models, it allows for category-specific adaptation, tailoring the model to the particular characteristics of each category, including the distribution of outcome labels. However, this approach does not account for potential correlations between the categories during estimation \citep[see e.g.][for a text classification of binary labels using BERT]{sun2019}. 

\subsubsection{Multi-task Classification}
Multi-task classification makes use of multi-task learning, where we keep the shared encoder model and add category-specific adapters for binary classifications. The encoder parameters $\theta$ are shared across the categories, and only the last hidden layer $h_i$ together with the output layer parameters $W$ and $b$ are optimized for the binary classification of the given category via an adapter \citep[compare e.g.][]{sun2019}. The computational demand is, however, reduced substantially compared to binary classification, as only one encoder model needs to be fine-tuned. On the downside, the full encoder parameters are not optimized for each given category.

\subsubsection{Multi-label Classification}
Multi-label classification frames the problem in a way that naturally fits our data, where a single review sentence can belong to multiple non-exclusive categories, i.e. $y_i = (y_{i,1},\dots, y_{i,C})$ is a binary vector with $y_{i,c}\in \{0,1\}$ for $C=12$ categories. In this approach, we fine-tune the model such that the number of output nodes matches the number of categories to predict \citep[see e.g.][for such multi-label classification approach]{Foster2024}. The key advantage is computational efficiency as only one model needs to be fine-tuned. In addition, the correlation among the categories is taken into account. In contrast, the problem of high imbalance in the labels becomes aggravated since many category combinations appear infrequently in the training data.\footnote{The annotated data contains 177 unique combinations of categories, of which 63 appear only once.}

\subsection{Performance Evaluation}

To evaluate the classification performance, we randomly divided the annotated set of 3,000 sentences into training (2,500) and test (500) sets while stratifying by the outcome category.\footnote{Due to the high class imbalance the stratification for the multi-label classification was not feasible.} We used the training set of 2,500 sentences to fine-tune the model and the untouched test set of 500 sentences to evaluate the prediction accuracy. To ensure a robust evaluation not dependent on a single data split, we performed five-fold cross-validation, stratified by outcome category, within the training set. Thus, we fine-tuned the model on four folds fo size 2,000 and evaluated it on the remaining fold of size 500, such that each fold served as a validation set once. We report the average performance metrics over the five folds (as recommended by \citet{hastie09}).

We calculated the following performance metrics: accuracy, precision, recall, and the F1 score \citep{rainio2024}. The accuracy is defined as the proportion of correctly classified texts:
\vspace{-0.1cm}
$$\text{Accuracy}= \frac{1}{n} \sum_{i=1}^n \mathbb{I}(\hat{y}_i^{\mathcal{C}} = y_i) ]$$
\vspace{-0.75cm}

\noindent while precision and recall measure the proportion of correctly classified texts of a given category out of all classified texts of the given category, and out of all actual texts of the given category, respectively:
\vspace{-0.1cm}
$$\text{Precision}= \frac{\sum_{i=1}^n \mathbb{I}(y_i=1 \wedge \hat{y}_i^{\mathcal{C}}=1)}{\sum_{i=1}^n \mathbb{I}( \hat{y}_i^{\mathcal{C}}=1)} \quad \text{and} \quad \text{Recall}= \frac{\sum_{i=1}^n \mathbb{I}(y_i=1 \wedge \hat{y}_i^{\mathcal{C}}=1)}{\sum_{i=1}^n \mathbb{I}(y_i=1)}.$$
\vspace{-1.0cm}

\noindent Our preferred evaluation measure is the F1 score, which is the harmonic mean of precision and recall:
\vspace{-0.1cm}
$$\text{F1}= \frac{2 \cdot \text{Precision} \cdot \text{Recall}}{\text{Precision} + \text{Recall}}$$

\noindent and takes the class distribution into account, i.e. the prevalence of the categories. Uneven proportions in the class labels pose a challenge for any classifier, as there are only a few examples to learn from (the class imbalance problem \citep{johnson2019}). Therefore, we not only evaluate the metrics for both labels combined, the so-called \textit{micro-average}, but also separately per label, i.e. in the subsets of $y_i=0$ and $y_i=1$, and compute the average of label-specific metrics, the so-called \textit{macro-average}. While micro-average is dominated by the more prevalent class label, macro-average weights both class labels equally and, as such, better reflects the overall classification quality, particularly in cases with high class imbalance. \citep{manning1999foundations}.

\section{Results}

\subsection{Prevalence of Categories in Human-Annotated Set}

Sentences could be assigned to no category or one or more categories. When aggregating annotations based on majority agreement, 18.1\% of sentences were not assigned to any of the twelve categories; 19.8\% fell into a single category, and most sentences fell into two (22.9\%) or three (22.7\%) categories, respectively. 16.4\% of the sample was classified into four or more categories (see Figure \ref{fig:distribution_appendix} in Appendix for details).

\begin{figure}[h!]
\caption{Prevalence of 12 categories in an annotated set of 3,000 sentences. \textit{Note}: Dark blue colors show the percentages of sentences classified into the same category by all annotators; light blue colors indicate sentences classified as the same category by two of three annotators.}
\label{fig:distribution_stacked}
\centering
\includegraphics[width=1\textwidth]{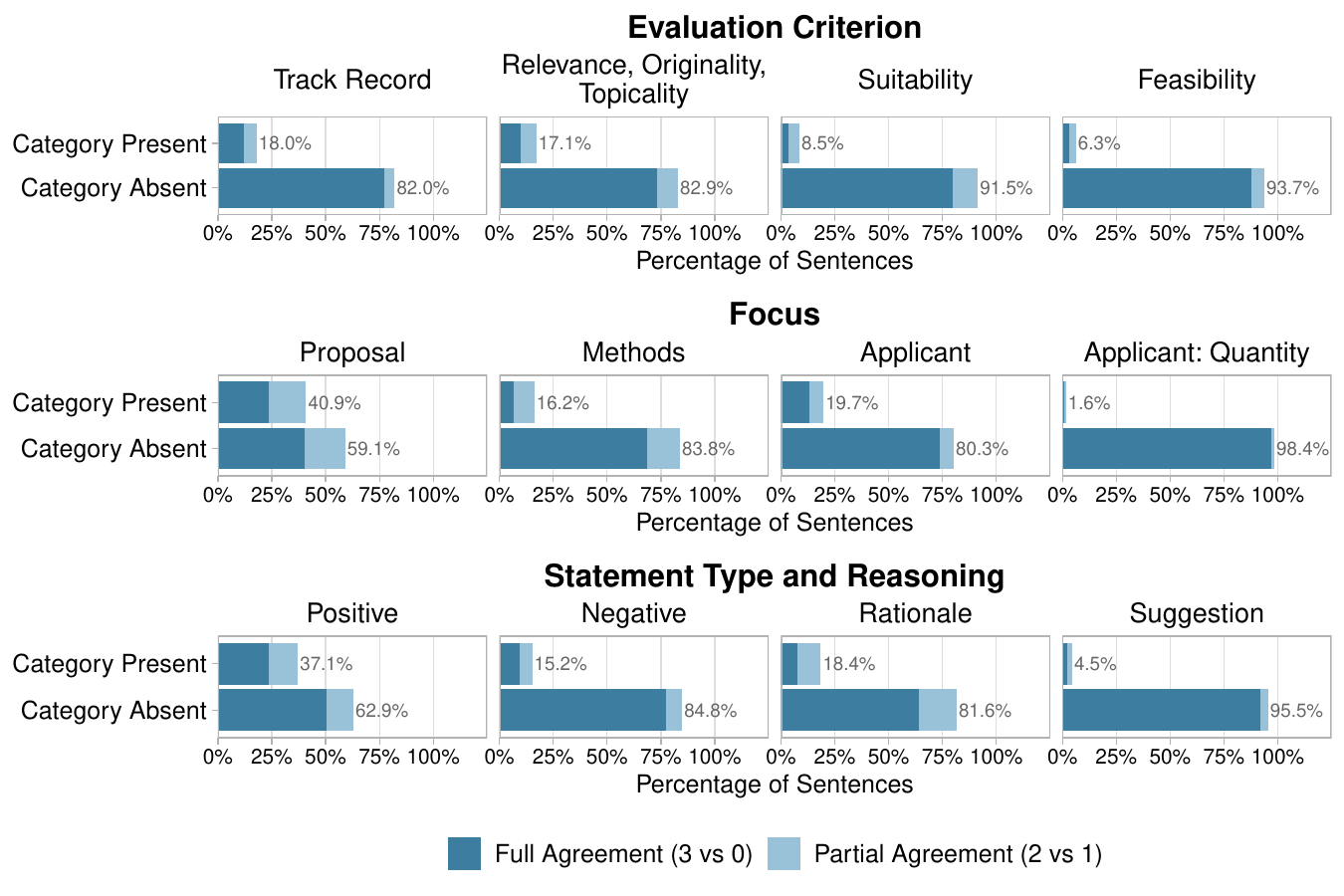}
\end{figure}

For each of the 12 categories, we first assessed the percentage of sentences where all three coders assigned the same label. Agreement ranged from 64\% (\textit{Proposal}) to 98\% (\textit{Applicant: Quantity}), with an average of about 80\% agreement across all categories.
Figure \ref{fig:distribution_stacked} provides a more detailed assessment of coding agreement and the prevalence of each category, revealing a considerable variation among them. Categories reflecting the \textit{Evaluation Criterion} aspect had prevalence rates ranging from   6\% to 8\% for  \textit{Suitability} and \textit{Feasibility}, respectively, and from 17\% to 18\% for \textit{Track Record} and \textit{Relevance, Originality, and Topicality}. For the categories addressing the \textit{Focus} aspect, a high prevalence was observed for \textit{Proposal} (41\%) \textit{Method} (16\%), and the \textit{Applicant} (20\%) in general, not considering the quantitative aspects. In contrast, only 1.6\% of the sentences mentions explicitly the quantitative aspects of the applicant (\textit{Applicant: Quantity}), such as the number of publications. The prevalence for the \textit{Statement Type and Reasoning} categories also varied substantially. The \textit{Positive} references are more prevalent (37.1\%) than \textit{Negative} ones (15.2\%), while not all of them provide a \textit{Rationale} for the statement.\footnote{The annotators identified a \textit{Rationale} only if they marked the sentence as either \textit{Positive} or \textit{Negative}. When considering only the sentences classified as \textit{Positive} or \textit{Negative} by at least two annotators, the proportion of sentences with a \textit{Rationale} (also agreed upon by at least two annotators) rises from 18.4\% to 35.5\%. Further details can be found in Appendix \ref{app:context} and Figure \ref{fig:rationale_context}.} Only 4.5\% of the 3,000 sentences provided suggestions.

\subsection{Comparison of Classifier Performance}

\subsubsection{Text Classification Approach}

We relied on the SPECTER2 model, pre-trained on scientific texts and augmented by citation graph, which we fine-tuned based on our set of annotated sentences. Table \ref{tab:f1_compare_all_main} compares the performance in terms of the F1 score for binary, multi-label and multi-task classification based on the test set of 500 review sentences.\footnote{Appendix contains detailed performance metrics for the three classification approaches (see Tables \ref{tab:f1_compare_detailed_binary}, \ref{tab:tab:f1_compare_detailed_multilabel}, and \ref{tab:tab:f1_compare_detailed_multitask}). We report the accuracy, balanced accuracy, class-specific F1 scores, macro-averaged F1 scores, micro-averaged F1 scores, precision, recall, and class-specific precision and recall. The main results always refer to macro-averaged F1 scores, which are not weighted by the class shares and thus provide a measure of classification performance that weights both class labels equally, as opposed to micro-averaged F1 scores.}  Fine-tuning with separate binary classifiers provided the best classification performance across the twelve categories. The macro F1 scores for the binary classifiers range from 0.71 (\textit{Rationale}) to 0.93 (\textit{Applicant: Quantity}), with an average of 0.85 across all twelve classifiers. The average F1 score for the multi-label approach was lower (0.73) and the range of F1 scores much larger (0.48 for \textit{Feasibility}; 0.90 for \textit{Track Record} and \textit{Applicant}). With an average F1 score of 0.62, the multi-task approach failed to produce reliable predictions. The performance of the models that share the information among the categories appears to suffer from the high-class imbalances in the category labels.

\input{tables/table_f1_compare_all_main.tex}

\begin{figure}[h!]
\caption{Macro F1 scores based on 5-fold cross-validation for all three classification approaches. Cross-validation folds consist of 500 sentences. Grey dots show the minimum and maximum F1 Scores across the 5 folds. Vertical blue dashed lines and numbers in bold show the average F1 scores across all twelve classifiers.}
\label{fig:f1_compare_all}
\centering
\includegraphics[width=1\textwidth]{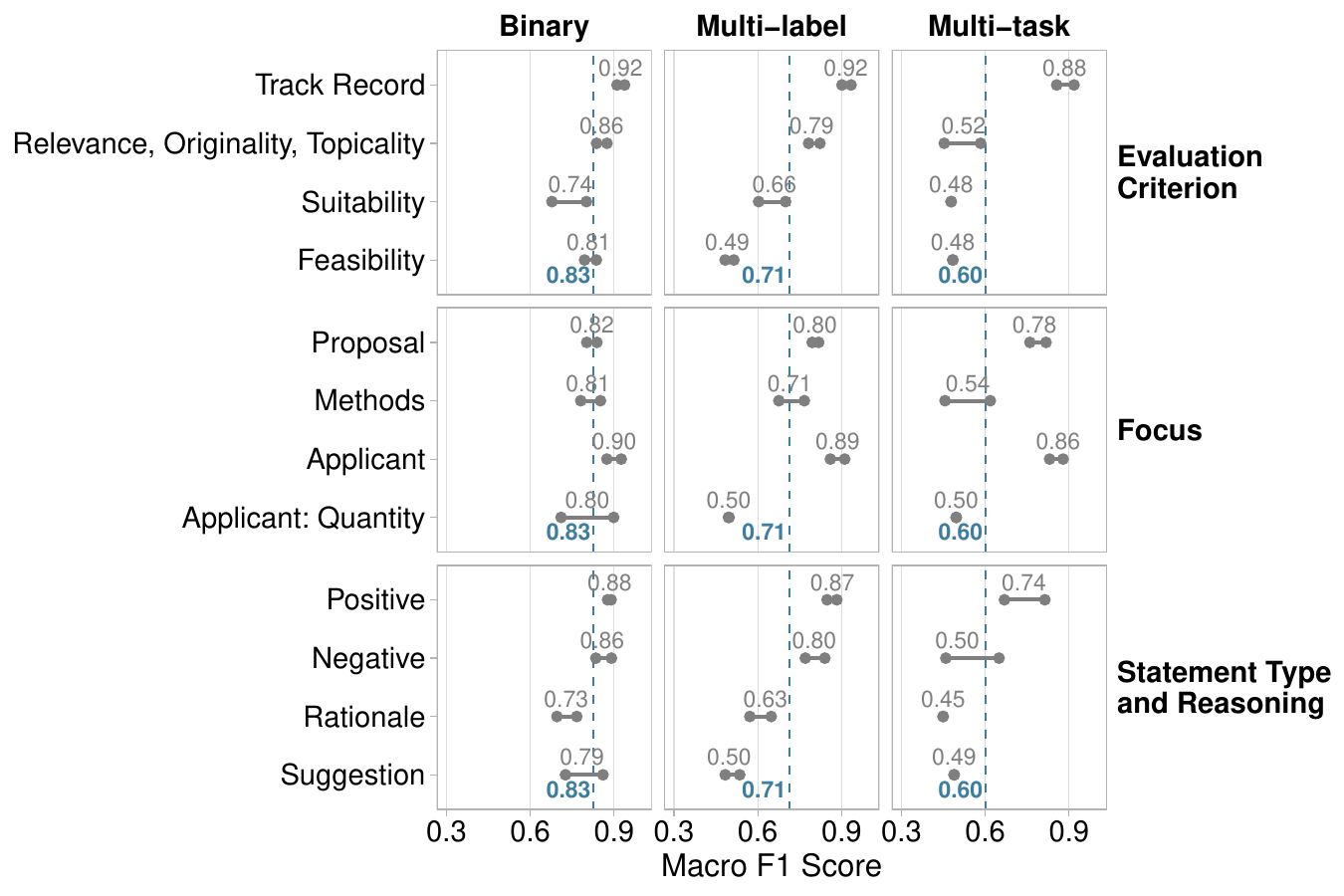}
\end{figure}

The fine-tuning of the models above relied on a single training-test split, as suggested by \citet{tunstall22}. In this case, the test set refers to a subset of annotated data reserved for evaluating the fine-tuned model's classification performance, ensuring an unbiased assessment of the model's performance. To assess whether the allocation of sentences into the test and training set might affect the overall conclusions, we conducted a five-fold cross-validation. Figure \ref{fig:f1_compare_all} compares the macro F1 score metrics for the five-fold cross-validation for the binary, multi-label, and multi-task classification. The dots show the range of the F1 scores across the five folds; the number above each range depicts the average F1 score across all folds. Mirroring the results from Table \ref{tab:f1_compare_all_main}, Figure \ref{fig:f1_compare_all} reveals that the binary classification approach outperforms the multi-label and multi-task classifiers. The average macro F1 score across all categories for the binary classifiers was 0.83, while the average F1 scores for multi-label (0.71) and multi-task (0.60) are lower. Moreover, the variation of F1 scores across categories is considerably smaller for the binary approach. The five-fold cross-validation further supports that optimal performance is achieved through binary classifiers specifically fine-tuned for each category. 

Overall, using separate binary classifiers for each category appears to be the most effective approach \citep[see also][]{severin23}, despite the greater computational efficiency of multi-label and multi-task methods. Therefore, we proceed with our analyses using the binary classification approach.

\subsubsection{Choice of Transformer Model}

Following the binary classification approach, we further assessed the suitability of the underlying pre-trained model and its impact on classification performance. Therefore, alongside SPECTER2, we also repeated fine-tuning for the baseline BERT model and the robustly optimized RoBERTa model. 

\input{tables/table_f1_compare_binary_pretrained.tex}

\begin{figure}[h!]
\caption{Cross-validation macro F1 scores for binary classifiers fine-tuned based on the SPECTER2 model, BERT, and RoBERTa. Circles show the minimum and maximum F1 scores; squares show the average F1 score. }
\label{fig:f1_compare_binary_pretrained}
\centering
\includegraphics[width=1\textwidth]{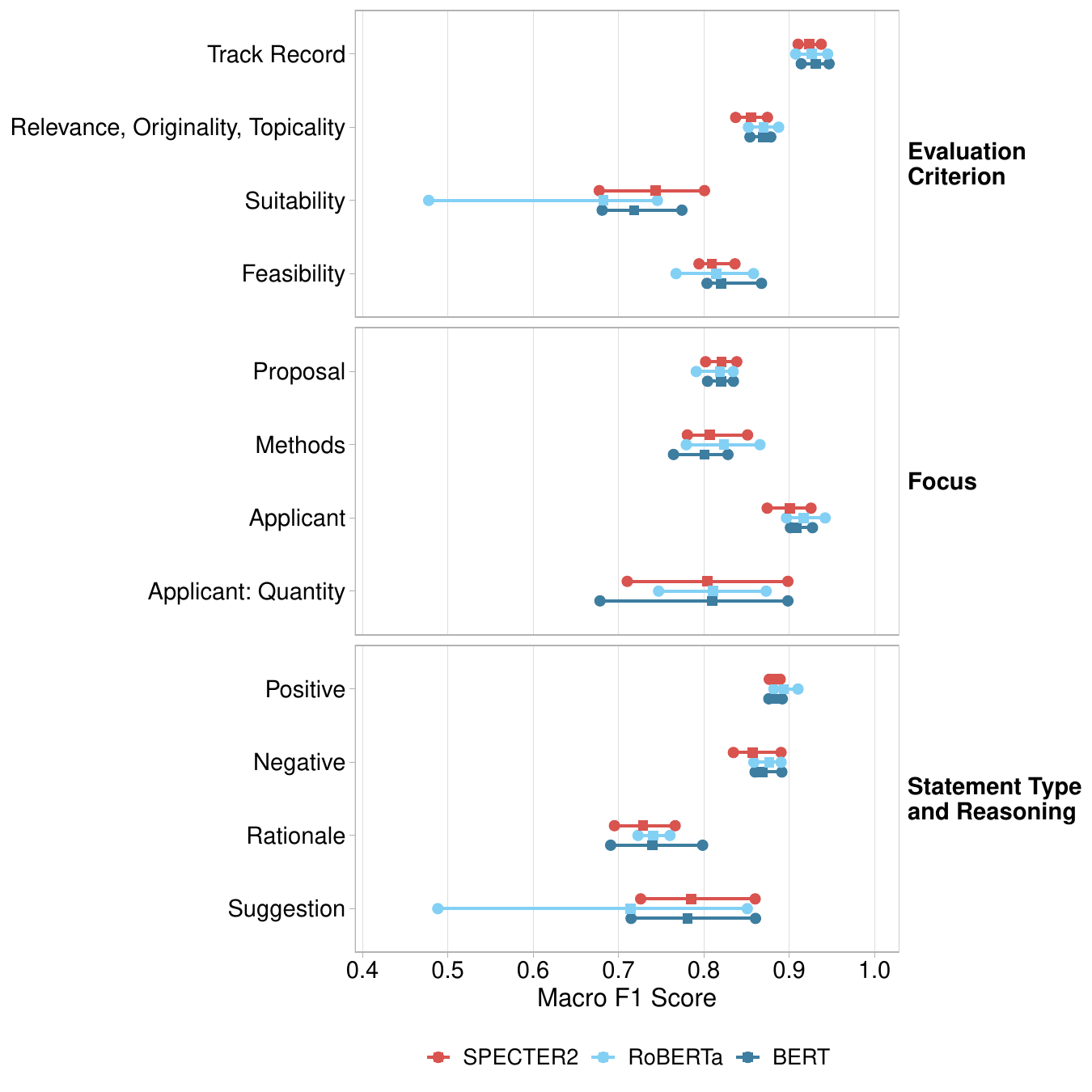}
\end{figure}

The results reveal similar performances of BERT and RoBERTa compared to the SPECTER2 model, both in terms of the test and cross-validation accuracy (see Table \ref{tab:tab:f1_compare_binary_pretrained}  and Figure \ref{fig:f1_compare_binary_pretrained}). Indeed, the average F1 score across all categories is 0.85 for all considered models in the test set. This result suggests that the annotated data used for fine-tuning is the decisive element rather than the model itself. This is also supported by the accuracy results provided in Table \ref{tab:tab:f1_compare_binary_pretrained}, which shows that almost all F1 scores were close to or above 0.8 except for the \textit{Rationale} category, which is inherently difficult to classify, even for human annotators. These accuracy levels exceed those found in related studies classifying peer review texts, despite much larger annotated sets \citep[see][]{ghosal22}.

\subsection{Relationships Between Classifier Performance, Prevalence, and Difficulty of Human-Annotation}

\begin{figure}[h!]
\caption{The relationship between the prevalence of categories and macro F1 scores for binary, multi-label, and multi-task classifiers. \textit{Note}: Pearson's correlation coefficients are shown in the bottom-right corners.}
\label{fig:f1_shares}
\centering
\includegraphics[width=1\textwidth]{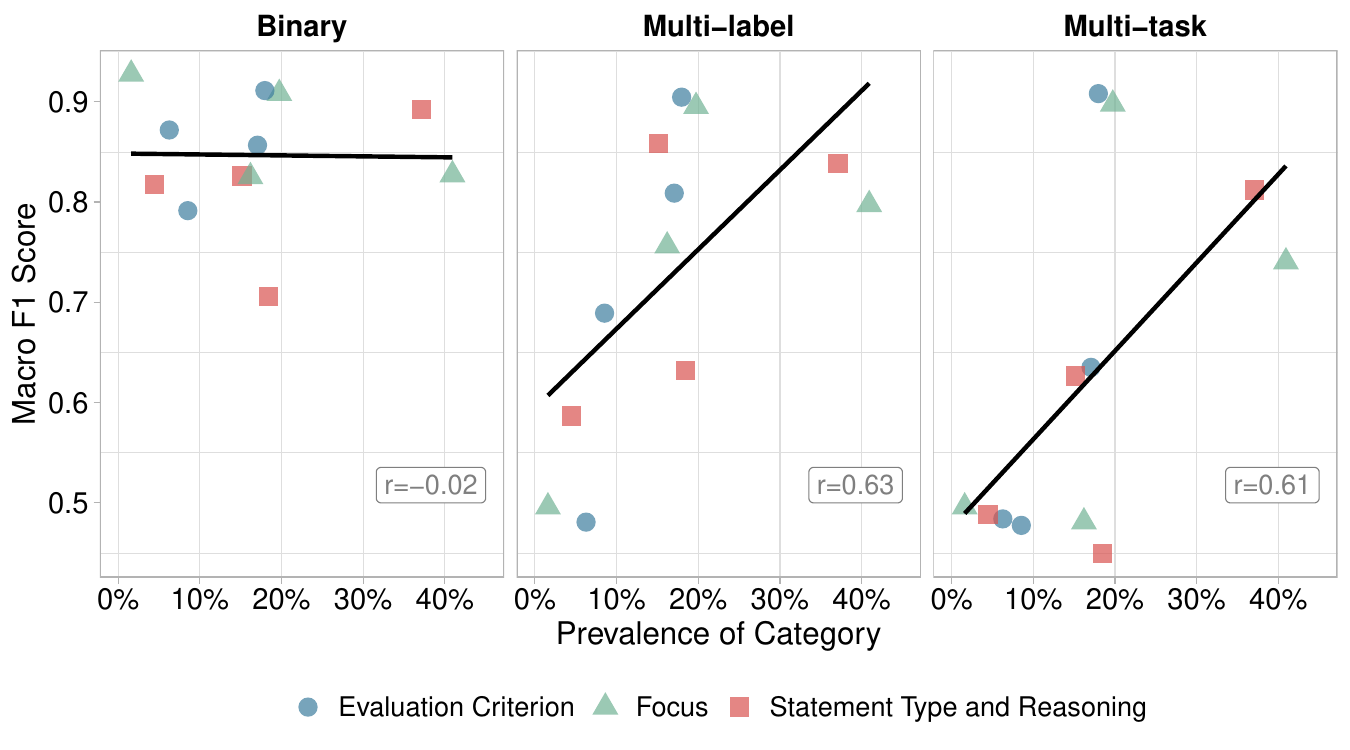}
\end{figure}

We examined variation in performance across classification methods and categories by relating performance metrics to the prevalence of each category in the training set and the findings from a survey distributed to the annotators.  Figure \ref{fig:f1_shares} shows the relationship between the prevalence of categories (in \%) and the F1 Score for the three classifiers based on Pearson's correlation coefficients. There were moderate to strong positive correlations between the F1 Score and the prevalence of the category for the multi-label and multi-task classifiers, but not for the binary classifier. This supports choosing the binary classifier, as its performance is unrelated to category prevalence or label imbalance.

We distributed a short questionnaire to the four annotators during the annotation process. It assessed annotators' perceived difficulty of identifying each of the twelve categories to better understand the annotators' confidence in their decisions \citep{han2023expert}.\footnote{The survey started with the following statement: \textit{``Reflecting on the previous coding rounds, we would like you to assess the ease or difficulty you experienced in classifying the categories featured in our coding tasks. Your personal perspective is highly valued, and we encourage you to share your honest opinion.''}} For each category, the survey asked: \textit{``On a scale ranging from 1 (very straightforward) to 10 (very challenging), how easy did you find classifying} \dots\textit{''} We collected the views of all four annotators and calculated the average score. 
\begin{figure}[h!]
\caption{The relationship between macro F1 scores for binary classifiers (left panel), prevalence of categories (right panel) and the perceived difficulty of annotating each category, based on survey distributed to the four annotators.  \textit{Note}: Pearson's correlation coefficients are shown in the top-left corners.}
\label{fig:f1_shares_diff}
\centering
\includegraphics[width=1\textwidth]{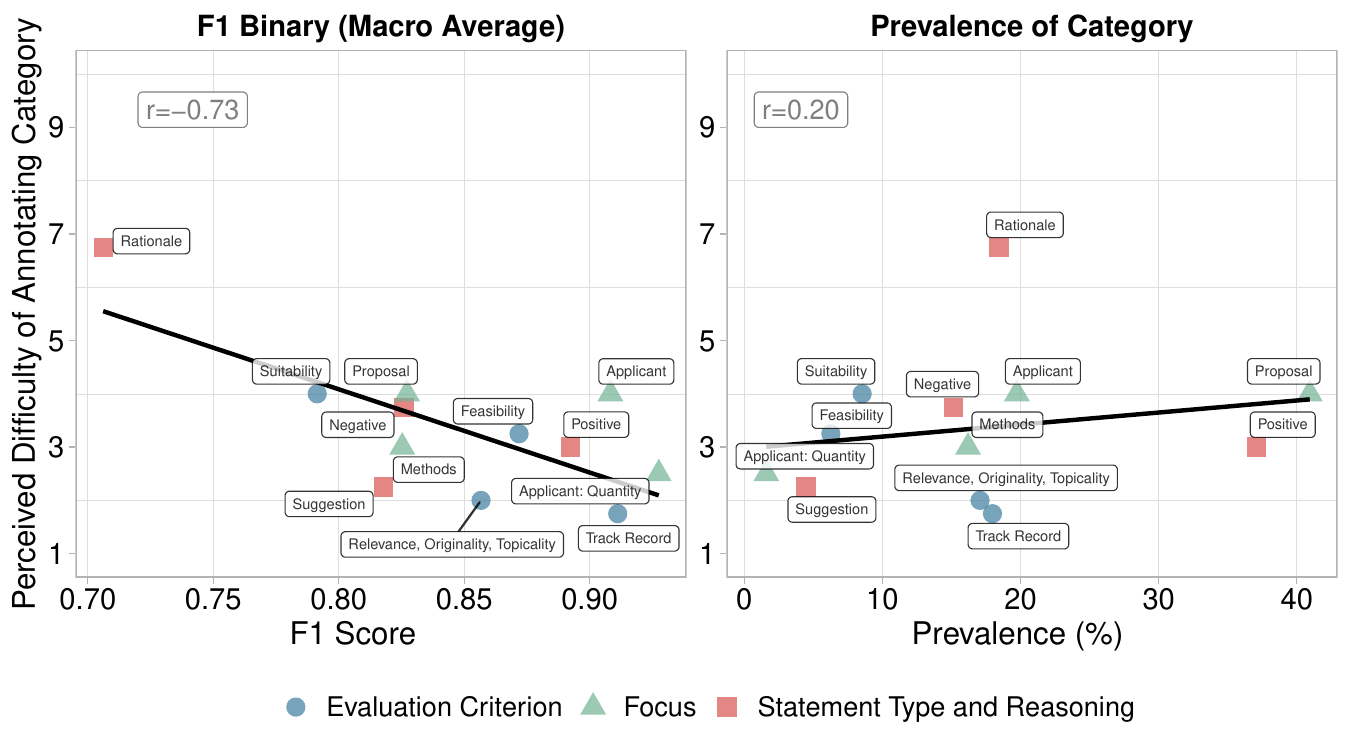}
\end{figure}

The survey reveals substantive differences in the perceived difficulty in identifying the categories. It ranged from 1.75 for \textit{Track Record} (fairly straightforward) to 7.25 for \textit{Rationale} (challenging). The average difficulty was 3.73; the median 3.50. Apart from \textit{Rationale}, the difficulty of all other categories was below 4.75. Assessing  \textit{Rationale} was the most challenging task.

Figure \ref{fig:f1_shares_diff} relates the perceived difficulty of annotating each category to the F1 Score (left panel) and the prevalence of the categories (right panel). Categories that annotators found easier to classify have higher F1 scores ($r=-0.73$). For instance, \textit{Rationale} was the most difficult to annotate and had the lowest F1 score, whereas the applicant's \textit{Track Record} was straightforward for annotators and classifiers. Overall, categories that are easier to annotate align with better classification performance. In contrast, the prevalence of a category only weakly correlated with the perceived difficulty  ($r=0.20$,  right panel of Figure \ref{fig:f1_shares_diff}, indicating that annotators did not perceive infrequent categories (such as \textit{Applicant: Quantity} or \textit{Suggestion}) as more difficult than frequent categories, such as \textit{Positive} or \textit{Proposal}.

\subsection{Explainability of Best-Performing Classifier}

We assessed the face validity of the classifiers' performance via explainable machine learning \citep{molnar20}. Following \citet{severin23}, we conducted a keyness analysis to identify terms that predict each category.

We applied the twelve binary classifiers to the full set of review reports consisting of 1,612,405 sentences. For each category, we compared words and phrases in sentences where the relevant review characteristic was present against sentences without it, based on model predictions. Multi-word expressions were automatically identified through collocation analysis. A keyness analysis then used chi-squared tests to evaluate each term and phrase between the two document groups \citep{benoit18}. Table \ref{tab:keyness} displayed the 25 terms and multi-word expressions with the highest keyness values for each category.

\pagebreak
\clearpage
\input{tables/keyness_fullsample.tex}
\pagebreak
\clearpage

The keyness analysis identifies specific terms associated with each category, providing insight into the content. Common terms in sentences classified under the \textit{Proposal} category include \texttt{project}, \texttt{proposal}, \texttt{topic}, and adjectives like \texttt{timely}, \texttt{ambitious}, and \texttt{topical}. Predictive terms for the \textit{Positive} category include \texttt{expertise}, \texttt{excellent}, \texttt{strong}, \texttt{timely}, and \texttt{outstanding}, while \texttt{lack}, \texttt{unclear}, \texttt{weakness}, and \texttt{concern} are common in the \textit{Negative} category. The term \texttt{unk}, a placeholder for the applicant and co-applicant names to maintain anonymity is the most predictive term for the \textit{Applicant} and \textit{Track Record} categories. Additionally, terms such as \texttt{published}, \texttt{h-index}, \texttt{citations}, \texttt{first\_author},  \texttt{google\_scholar}, and \texttt{scopus} show high keyness values for \textit{Applicant: Quantity}.

\begin{figure}[h!]
\caption{Correspondence between prevalence in human-annotated set of sentences and predictions for sentences from review reports submitted between October 2016 and April 2024.}
\label{fig:correlation_prev}
\centering
\includegraphics[width=01\textwidth]{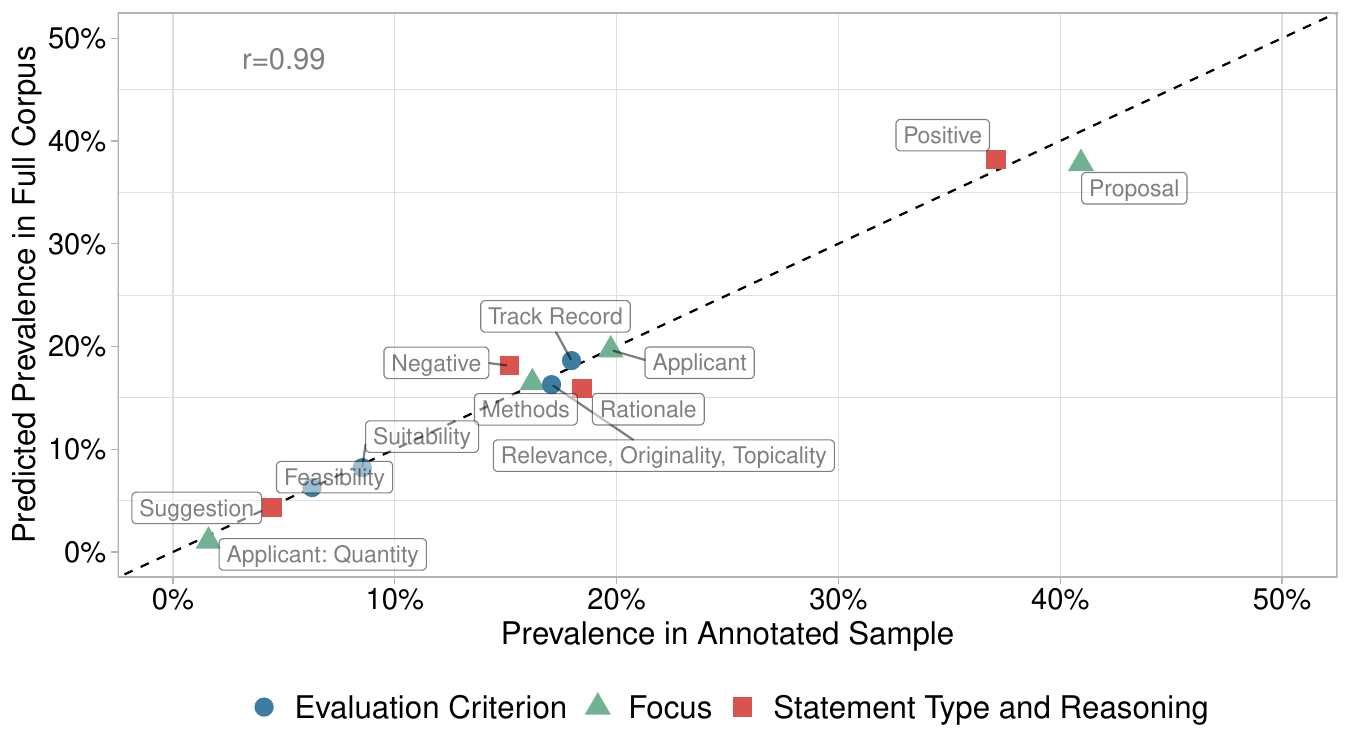}
\end{figure}

Further, we compared the prevalence of categories in the annotated sample of 3,000 sentences with the predicted prevalence in Project Funding review sentences from 2016 to 2024. A reliable classification should show correspondence between these aggregated percentages \citep{muellerfujimura}. Figure \ref{fig:correlation_prev} confirms that aggregated predictions for 1,612,405 review sentences align with the proportions in the annotated sample of 3,000 sentences ($r=0.99$). 

\subsection{The Prevalence of Categories in Peer Review Reports}

In order to gain substantive insights into the content of the grant peer review reports, we compare the prevalence of each category at the level of reviews submitted between 2016 and 2024. For each review and category, we measure prevalence by dividing the number of sentences classified as belonging to the given category by the total number of review sentences. A value of 30\% implies that three out of ten sentences address a given category. As sentences often contain more than one category, the prevalence across the 12 categories can exceed 100\%.

\begin{figure}[h!]
\caption{Distribution of the predicted prevalences across all sentences in each review, for each category. Vertical dashed lines show the average prevalence across the 47,522 review reports.}
\label{fig:prevalence_histogram}
\centering
\includegraphics[width=01\textwidth]{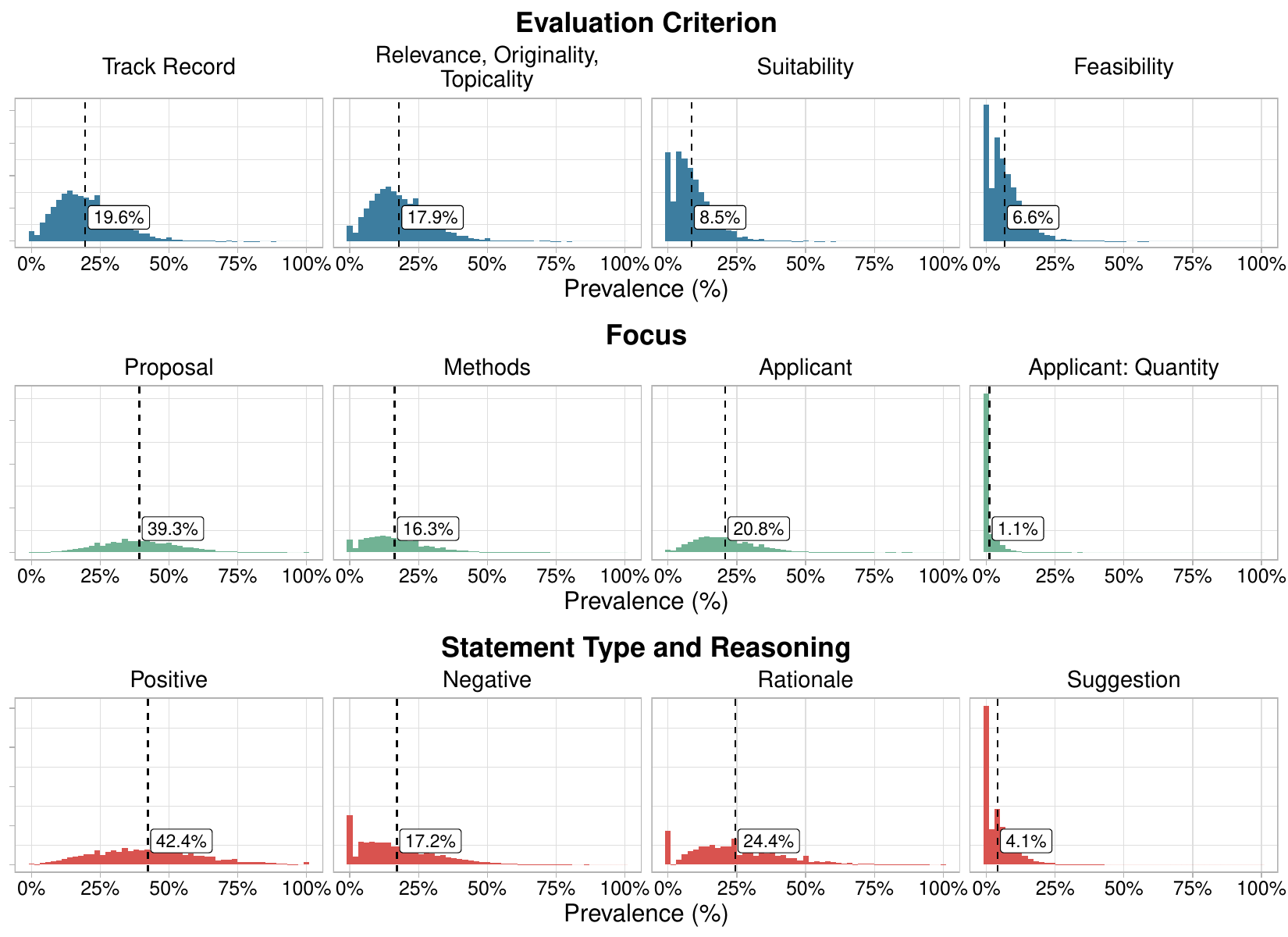}
\end{figure}

Figure \ref{fig:prevalence_histogram} reveals considerable variation across reviews and categories. In the \textit{Evaluation Criterion} dimension, the \textit{Track Record} of an applicant is most prevalent (19.6\% on average across all reviews), followed by \textit{Relevance, Originality, Topicality} (17.5\%), whereas \textit{Suitability} (8.5\%) and \textit{Feasibility} (6.6\%) are less prevalent. The highest \textit{Focus} of the reviews is on the \textit{Proposal} (39.3\%), while the focus on methods is lower (16.3\%). More sentences mention the \textit{Applicant} in general (20.8\%) in comparison to small percentage addressing quantity-related aspects of the applicant (1.1\%). Turning to the \textit{Statement Type and Reasoning}, we observe that \textit{Positive} sentences outweigh \textit{Negative} sentences (42.4\% vs. 17.2\%). The average review provides a \textit{Rationale} in 24.4\% of the sentences that are classified as positive or negative.\footnote{Since \textit{Rationale} is only considered for \textit{Positive} or \textit{Negative} statements in the annotation, we aggregate the predictions for \textit{Rationale} category as follows: if a sentence's prediction is either \textit{Positive} or \textit{Negative}, we retain its \textit{Rationale} prediction. If the sentence's prediction is neither \textit{Positive} nor \textit{Negative}, we disregard its \textit{Rationale} prediction. Subsequently, we calculate the review-level prevalence by considering only the sentences classified as \textit{Positive} or \textit{Negative} in the denominator of the average.} \textit{Suggestions} are infrequent in grant peer review reports (4.1\%).

While the averages provide a measure of central tendency, the histograms reveal considerable variation across the reports. For example, while the prevalence of \textit{Positive} sentences is symmetrically distributed around the mean, \textit{Negative} sentences are skewed, with 10,157 reviews containing fewer than 5\% of negative sentences, while only 946 reviews include more than 50\% of negative review sentences.

\subsection{Robustness Tests and Validation}

We performed several robustness tests and validation analyses.

\subsubsection{The Value of Additional Training Data}\label{sec:ablation}

During annotation, it is crucial to determine the amount of labeled data needed to train classifiers with sufficient accuracy. Since estimating the minimum required annotated texts is challenging \textit{a priori}, we use an approximation method -- a so-called ablation study -- to assess the added value of additional annotated samples for classification \citep{Foster2024, bucher2024}. We split the annotated sample of 3,000 sentences into a training set of 2,500 sentences and a test set of 500. We further divide the training set into chunks of 500 sentences. Then, we perform \textit{sequential fine-tuning} on the pre-trained SPECTER2 model, gradually adding chunks of 500 sentences to the training set, up to 2,500 sentences. At each step, we evaluate classification accuracy on the test set. 

\begin{figure}[h!]
\caption{F1 scores (macro-average) depending on the size of the training set}
\label{fig: performance_metrics_ablation}
\centering
\includegraphics[width=0.99\textwidth]{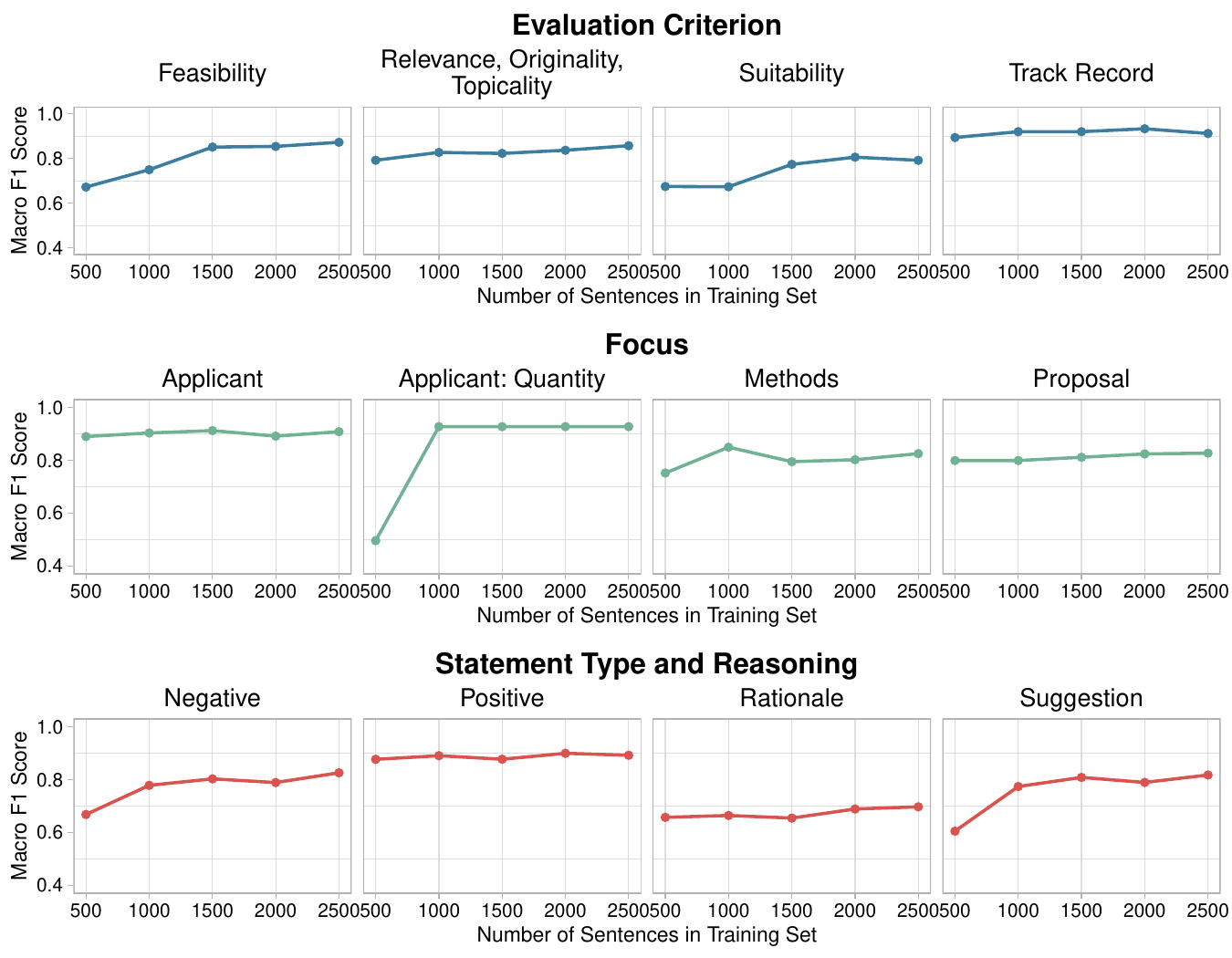}
\end{figure}

Figure \ref{fig: performance_metrics_ablation} reveals two main findings. First, for well-balanced categories like \textit{Positive} and those with moderately imbalanced classes like \textit{Applicant}, high classification accuracy was achieved even with a small sample of annotated data, and additional data only marginally increased accuracy. Second, adding training sentences substantially improved classification accuracy for more imbalanced categories, such as \textit{Feasibility} or \textit{Suggestion}. Since the greatest accuracy improvements came from the highly imbalanced categories, additional training data is particularly valuable for underrepresented classes. However, even for these imbalanced classes, accuracy gains become minimal once around 2,000 annotated sentences are used \citep[see][for similar findings]{Foster2024, bucher2024}. Given this evidence and the cost of obtaining further training data, we stopped additional annotations at 2,500 sentences in the training set.

\subsubsection{Longer Context Evaluation}

Although the sentence level is a common unit of analysis for text classification \citep[see e.g.][]{ghosal22,severin23}, we also explored whether providing a larger context would improve classification accuracy. Specifically, we focused on \textit{Rationale}, as its classification accuracy was significantly lower than that of other categories and often required more than one sentence to convey its full meaning. We additionally annotated  \textit{Rationale} to capture if the rationale is provided in the surrounding sentences, i.e., one sentence before and after the target sentence. We then defined a new category, \textit{Rationale and Context}, combining annotations for the target sentence with those of surrounding sentences (see Figure \ref{fig:rationale_context} in Appendix \ref{app:context} for the prevalence statistics). Finally, we repeated the fine-tuning step for this category by providing the target sentence and the surrounding sentences as input to the transformer model.

The classification accuracies were comparable with an F1 score of 0.71 for both \textit{Rationale} and \textit{Rationale and Context} (Table  \ref{tab:tab:f1_compare_rationale} in Appendix \ref{app:context}).  Figure \ref{fig:f1_compare_binary_rationale} provides further evidence based on five-fold cross-validation, suggesting that the model fine-tuned with only the target sentence tends to outperform the context-inclusive model. Adding surrounding sentences does not improve the classification accuracy for \textit{Rationale}.

\subsubsection{Limiting the Training Set to Sentences with Full Agreement}

To evaluate how the composition of training data affected classification performance, we compared classifiers trained on the complete set of sentences with majority annotator agreement to those trained exclusively on sentences with unanimous agreement. Classifiers trained on full-agreement data performed slightly better in categories with a lower inter-coder agreement, such as \textit{Proposal} and \textit{Positive}, while categories like \textit{Applicant: Quantity} and \textit{Feasibility} saw a notable decline in accuracy (Table \ref{tab:tab:f1_compare_binary_full} in Appendix \ref{trainingvsfull}). To further examine this, we conducted a cross-validation exercise where training folds included only full-agreement sentences, while validation folds covered all 500 sentences. The results indicated that restricting the training set to full-agreement sentences did not significantly improve overall classification performance. Therefore, using a larger training set based on majority voting seems advantageous in balancing training data quality and quantity.

\subsubsection{Additional Training Data Through Few-Shot LLM Classification}

Using few-shot learning, Large Language Models (LLMs) can be employed to annotate additional data, which are then used to fine-tune smaller transformer models \citep{Pelaez2024,Gupta2024}, or directly used for text classification \citep{brown2020language}. We explored the potential of generating additional training data using the Llama model \citep{touvron2023llama}, in particular the \textsf{Meta-Llama-3-8B-Instruct} version \citep{llama3modelcard}. Specifically, we tested the model's ability to classify the twelve categories by developing tailored prompts based on the Hugging Face chat template.\footnote{\url{https://huggingface.co/docs/transformers/main/chat_templating}.} As part of the prompt, we provided one example of a sentence representing a given category and one example of a sentence not representing any category. A detailed prompt structure is depicted in Figure \ref{fig:chat_template} in Appendix.

We evaluated the classification performance based on the same set of 500 sentences used in the test set in the main classification analysis. The LLM performed poorly, particularly in classifying sentences for imbalanced categories, leading to low F1 scores. The results suggest that LLM annotation does not achieve the accuracy of human annotation and lags behind the accuracy of fine-tuned BERT models (see Table \ref{tab:llama_annotation}) as has been previously documented by \citet{sun2023text} or \citet{bucher2024}. Consequently, we neither include artificially generated data in our models nor use LLMs for text classification directly.

\section{Discussion}
Peer review is a fundamental component of the research lifecycle, serving as a quality control process where experts evaluate submissions for publication, promotion or funding. Although peer review is a central element in evaluating grant proposals, research into the content of grant peer review reports is scarce. Such research could examine the quality of submitted reports and thus address the decreasing trust in the review process \citep{langfeldt24}, promote fairness and consistency in panel discussions, increase process transparency, and identify areas for reviewer training. 

The tools used for such tasks must be carefully developed, tested and transparently described. In this study, we have presented a pipeline for developing classifiers to aid in analyzing the contents of grant peer review reports. Our results show that machine learning can contribute to the assessment of review texts, but some characteristics are analyzed more reliably than others. In addition to the fine-tuned classifiers, which we share open-source, the step-wise procedure described in this paper may enable other groups to develop their own peer review text classification approaches.

Assessment of peer review reports using machine learning methods has gained considerable attention in recent years \citep{sizo19}. Building on existing work \citep{Hua2019, Fromm2021, ghosal22,  hren2022, GuoetAl2023, severin23, williams23}, we established and validated a classification pipeline by defining review characteristics relevant to funding organizations, creating an annotation codebook, manually annotating a set of review sentences, and fine-tuning pre-trained transformer models, thus enabling the classification of reviews at scale. Furthermore, we compared various methods and approaches to test the robustness of our findings. Our approach offers a reliable framework for systematically assessing and potentially improving peer review content by employing a rigorous annotation process and transformer-based classifiers. 

Developing an effective annotation codebook is a critical process, requiring careful refinement to ensure accuracy in identifying key characteristics of peer review reports in line with the funder's requirements   \citep{hug20}. 
Annotators in this study were academics familiar with the peer review process, though not domain experts. It has been observed that instructed non-expert annotators typically adhere more closely to the guidelines in the annotation codebook and demonstrate performance comparable to that of expert annotators \citep{snow2008cheap}. Annotators must navigate technical jargon, implicit meanings, and vague and unclear texts. Working in a team of annotators can partly overcome these challenges. The quality of human annotations is crucial, as machine learning classifiers can only perform as well as the data they are trained on \citep{han2023expert}. When human annotators find specific categories challenging to classify, machine learning models, even advanced ones, will also struggle. 

As reflected in the lower inter-coder agreement, categories that proved inherently difficult to annotate include categories like \textit{Proposal} and \textit{Rationale}.  Narrow and clearly defined categories, such as \textit{Applicant: Quantity} and \textit{Suggestion}, showed high levels of agreement among annotators.  In contrast, broader categories allow more room for interpretation. The category \textit{Proposal}  is particularly broad, encompassing sentences that may address various sections of the grant proposal, including the research plan, descriptions of the proposed topic, or the literature review. Annotators also reported difficulties distinguishing this category from references made to proposed methods, especially in the context of social sciences and humanities.

For the category \textit{Rationale}, disagreements often arose from differing interpretations of what constitutes a satisfactory justification for a positive or negative statement. This subjective aspect was unavoidable and may reflect similar challenges that reviewers may encounter when articulating their reasons for a supporting or a critical statement. Another challenge affecting classification accuracy relates to class imbalance within the training data, i.e. low shares of specific categories, as machine learning models require a balanced set of examples for effective learning. Although LLMs offer the potential for handling imbalanced categories by generating additional labeled data (e.g. \citet{Gupta2024}; \citet{Pelaez2024}), human annotation remains superior when dealing with rare or nuanced categories.

Selecting the appropriate classification approach is also crucial, as different methods offer varying levels of performance depending on the complexity of the task. In this study, binary classification outperformed multi-label and multi-task approaches. The classification accuracy achieved shows F1 scores that align with findings from recent studies, such as \citet{severin23}, and even surpass those from other analyses like \citet{ghosal22}. Notably, models trained on domain-specific data did not significantly outperform more general models, as also documented by  \citet{Foster2024}, suggesting that comparing different modelling approaches can be more effective than over-optimizing the model choice. The choice of the unit of analysis can also influence performance, and sentence-level analysis employed in similar studies  \citep{ghosal22, severin23} proved more effective than classifying longer sequences in this context. The similar performance of the different models might indeed stem from the unit of analysis being a single sentence, where the additional value of pre-training on the scientific texts is potentially limited. Finally, researchers should select classifiers based on quantitative metrics, such as accuracy thresholds, and also qualitative assessments, like identifying systematic deviations from human annotations, to ensure trustworthy substantive analyses. Importantly, the classifiers should not be used for automatic classification of grant peer review reports without human oversight and the level of the classification accuracy should be critically evaluated before deployment.

Beyond findings related to the annotation and classification process, this study offers insights into the characteristics and prevalence of elements in grant peer review reports submitted to a national research funder \citep{hug2024referees}. The peer review reports predominantly focused on the proposal and methods rather than the applicant. This finding aligns with the funding scheme for which the review reports were written: SNSF Project Funding is a project- rather than a person-centered funding scheme. It is reassuring that quantitative assessments of applicants based on the number of publications or bibliometric indicators such as citation counts, the h-index or impact factors were uncommon, in line with the Declaration of Research Assessment (DORA, see \citet{bladek2014dora}) signed by the SNSF in 2014. It will be interesting to compare peer review reports for project-centered funding scheme with reports submitted to person-centered funding schemes that promote early career researchers. Further, the findings reveal that grant peer reviews tend to be more positive than negative, a result in line with previous studies of grant proposals  \citep{hren2022} but contrary to evidence from journal peer review  \citep{severin23}. While over 30\% of sentences from a typical journal peer review contained suggestions \citep{severin23}, the average predicted prevalence of suggestions in our sample of grant peer reviews was only 4\%. Notably, most of the positive or negative statements (67\%) in the subset of 3,000 annotated sentences came with a rationale for the statement, i.e. justification for the praise or criticism was provided in the target sentence or the surrounding sentences (see Figure \ref{fig:rationale_context}). Overall, these first descriptive findings offer valuable insights into grant peer review practices, illustrating the potential of further analyzing whether reviewers adhere to the scheme's objectives and funders' guidelines. Detailed analyses are ongoing and will be the subject of future articles.

In conclusion, the pipeline developed in this study lays the groundwork for more detailed and systematic assessments of grant peer review reports. It enables research funding agencies to systematically evaluate and refine practices, with the ultimate aim of improving the evaluation process. Researchers may benefit from a clearer and more structured understanding of the peer review process, which could lead to increased engagement and adherence to review guidelines.

\pagebreak

\section*{Data Availability}
Due to data privacy laws, no raw data from this paper can be publicly shared. For a detailed description of data protection please refer to the data management plan underlying this work: \url{https://doi.org/10.46446/DMP-peer-review-assessment-ML}. The annotation codebook developed in this work is available online: \url{https://doi.org/10.46446/Codebook-peer-review-assessment-ML}.

\section*{Code Availability}
The code scripts for fine-tuning the classification models and the models themselves resulting from this work are available open-source on GitHub: \url{https://github.com/snsf-data/ml-peer-review-analysis}, archived on Zenodo: \url{https://doi.org/10.5281/zenodo.14215058} and Hugging Face Hub: \url{https://huggingface.co/snsf-data}, archived on Zenodo as well: \url{https://doi.org/10.5281/zenodo.14217855}.

\section*{Competing Interests}
This work was financed by the Swiss National Science Foundation (SNSF). Matthias Egger, Katrin Milzow, Anne Jorstad, Michaela Strinzel and Gabriel Okasa are employed by the SNSF.

\pagebreak
\singlespacing
\bibliographystyle{apalike}
\bibliography{references.bib}

\doublespacing

\clearpage

\appendix

\pagenumbering{arabic}
  \setcounter{figure}{0}
  \renewcommand{\thefigure}{A\arabic{figure}}
  \setcounter{table}{0}
  \renewcommand{\thetable}{A\arabic{table}}
  \setcounter{footnote}{0}

  \pagenumbering{arabic}
  \renewcommand*{\thepage}{A\arabic{page}}
  
\section{SNSF Project Funding and the Review Process}

\begin{figure}[h!]
\caption{Overview of the SNSF evaluation procedure} \label{fig:evaluation_proc}
\centering
\includegraphics[width=1\textwidth]{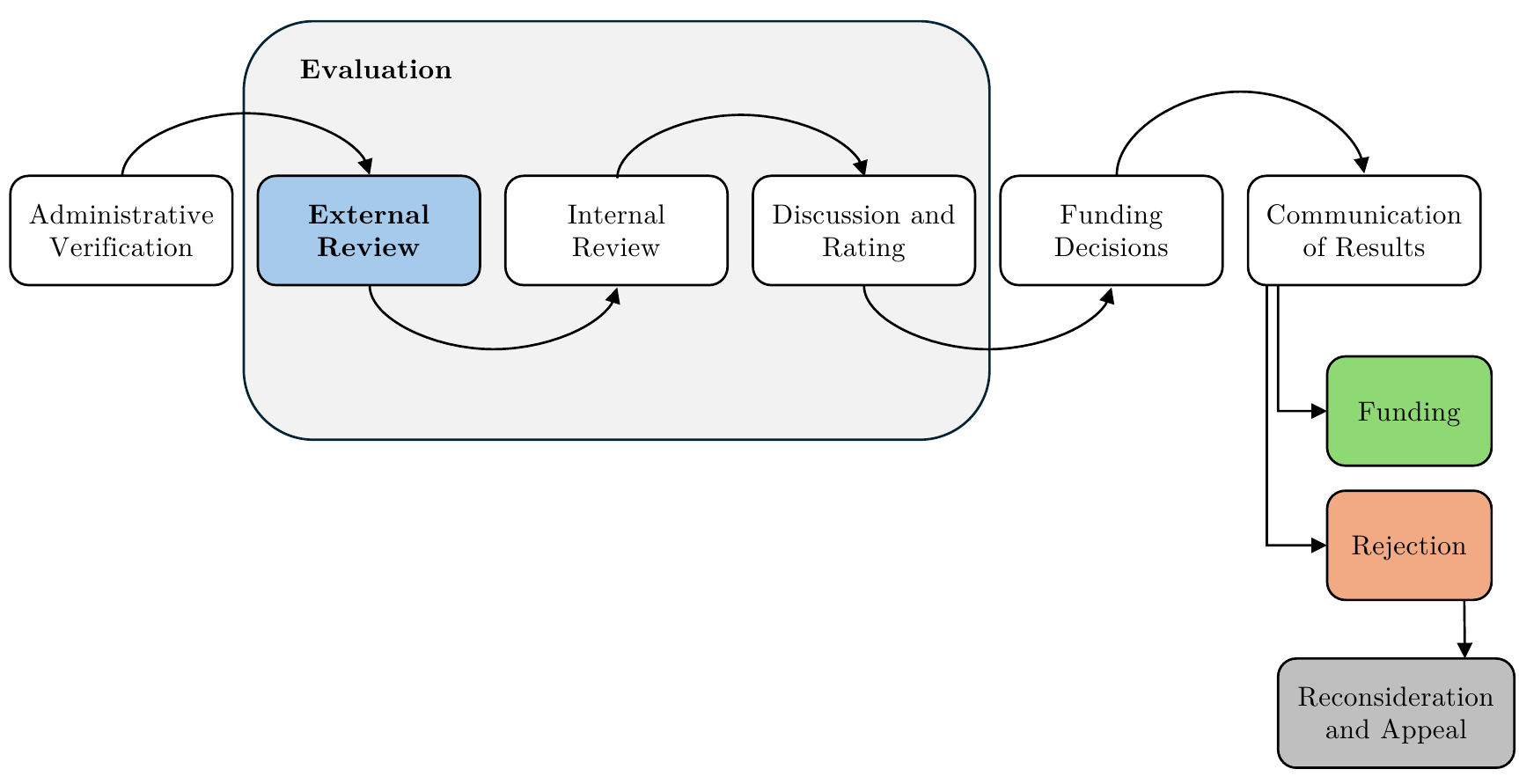}
\end{figure}

\pagebreak

\begin{figure}[h!]
\caption{Evaluation criteria for SNSF project funding applications}\label{fig:evaluation_proc_main}
\centering
\includegraphics[width=0.8\textwidth]{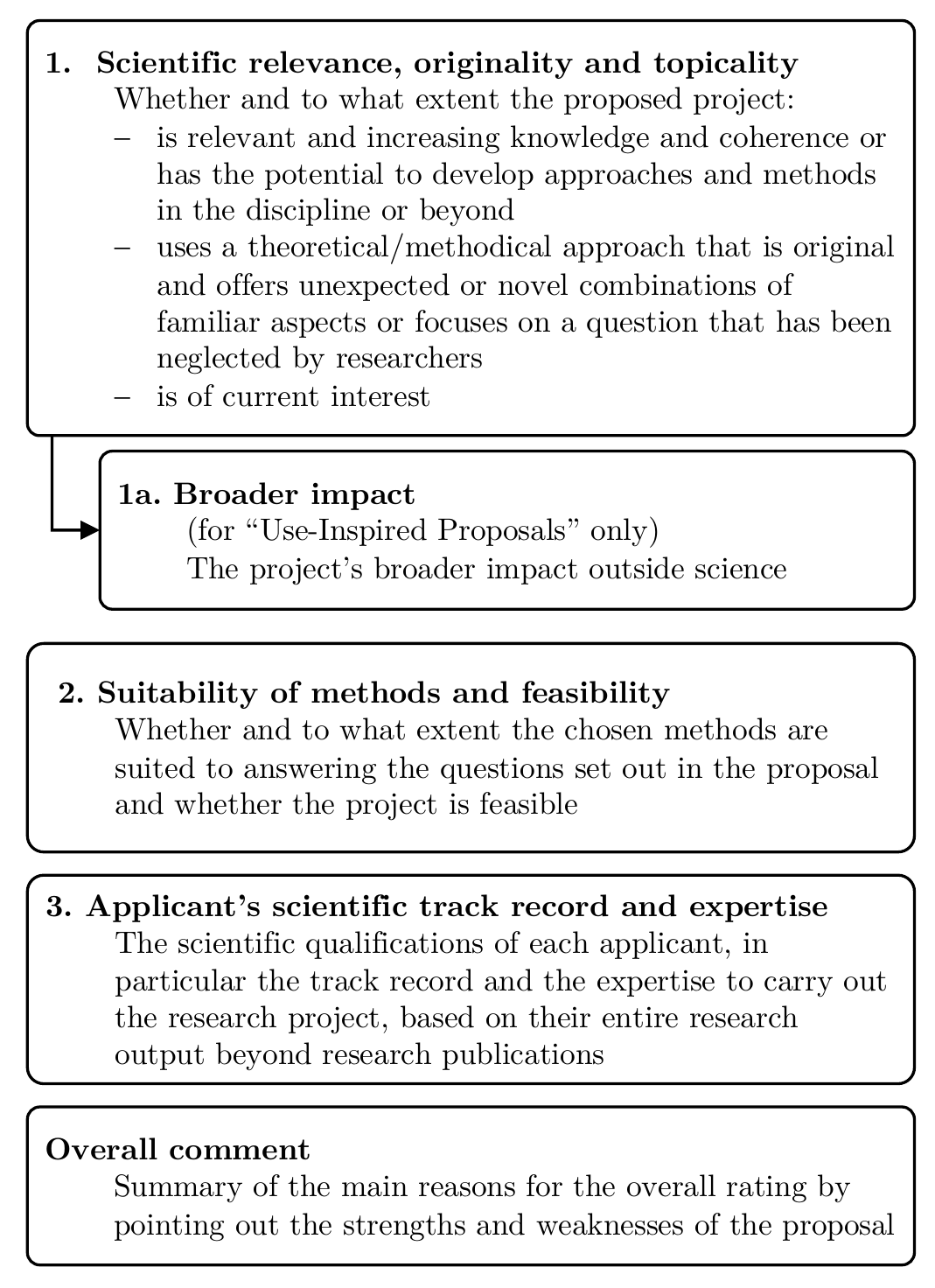}
\end{figure}

\pagebreak

\section{Overview of Coding Rounds}

Table \ref{tab:appendix_overviewrounds} provides an overview of all annotation rounds, the categories considered, and details on the sampling procedure for each round. Note that some of the categories, such as \textsf{impact\_beyond} (\textit{Impact beyond Academia}) and \textsf{implicit\_evaluation} (\textit{Implicit Evaluation}), have been trialed in some of the annotation rounds but were not used in the three main annotation rounds used for the analysis in this manuscript.

\begin{longtable}{p{0.06\textwidth} p{0.11\textwidth} p{0.11\textwidth} p{0.4\textwidth} p{0.2\textwidth}}
\caption{Overview of Annotation Rounds, Numbers of Annotators, and Sample Size} \label{tab:appendix_overviewrounds} \\ 
\hline
Round & Annotators & Sentences & Categories & Notes/Details \\ 
\hline
\endfirsthead

\caption[]{Overview of Coding Rounds, Numbers of Coders, and Sample Size (continued)} \\
\hline
Round & Annotators & Sentences & Categories & Notes/Details \\ 
\hline
\endhead

\hline
\endfoot

\hline
\endlastfoot

\#1 & 2 & 200 & addresses\_criterion, weakness, strength, segmentation & Stratified sample of 50 sentences for each of the four evaluation criteria \\ \hline
\#2 & 2 & 100 & addresses\_criterion, weakness, strength, rationale, suggestion, candidate, proposal, segmentation & Stratified sample of 25 sentences for each of the four evaluation criteria \\ \hline
\#3 & 2 & 100 & addresses\_criterion, weakness, strength, rationale, suggestion, candidate, proposal, segmentation & Stratified sample of 25 sentences for each of the four evaluation criteria \\ \hline
\#4 & 2 & 100 & criterion\_track\_record, criterion\_relevance\_originality\_topicality, criterion\_feasibility, criterion\_suitability, candidate\_other, candidate\_quantity, proposal\_general, proposal\_method, positive, negative, suggestion, rationale & Stratified sample of 25 sentences for each of the four evaluation criteria \\ \hline
\#5 & 2 & 100 & criterion\_track\_record, criterion\_relevance\_originality\_topicality, criterion\_feasibility, criterion\_suitability, candidate\_other, candidate\_quantity, proposal\_general, proposal\_method, positive, negative, suggestion, rationale, rationale\_context & Stratified sample of 25 sentences for each of the four evaluation criteria \\ \hline
\#6 & 4 & 48 & criterion\_track\_record, criterion\_relevance\_originality\_topicality, criterion\_feasibility, criterion\_suitability, candidate\_other, candidate\_quantity, proposal\_general, proposal\_method, positive, negative, suggestion, rationale, rationale\_context & Stratified sample of 12 sentences for each of the four evaluation criteria \\ \hline
\#7 & 4 & 48 & criterion\_track\_record, criterion\_relevance\_originality\_topicality, criterion\_feasibility, criterion\_suitability, candidate\_other, candidate\_quantity, proposal\_general, proposal\_method, positive, negative, suggestion, rationale, rationale\_context & Stratified sample of 12 sentences for each of the four evaluation criteria \\ \hline
\#8 & 4 & 1,000 & criterion\_track\_record, criterion\_relevance\_originality\_topicality, criterion\_feasibility, criterion\_suitability, applicant\_other, applicant\_quantity, proposal\_general, proposal\_method, positive, negative, suggestion, rationale, rationale\_context, implicit\_evaluation, impact\_beyond & Random sample of entire corpus; no stratification based on evaluation criteria; English sentences only \\ \hline
\#9 & 4 & 1,000 & criterion\_track\_record, criterion\_relevance\_originality\_topicality, criterion\_feasibility, criterion\_suitability, candidate\_other, candidate\_quantity, proposal\_general, proposal\_method, positive, negative, suggestion, rationale, rationale\_context, impact\_beyond & Random sample of entire corpus; no stratification based on evaluation criteria; English sentences only \\ \hline
\#10 & 4 & 1,000 & impact\_beyond & Selection based on probabilities returned by fine-tuned classifier \\ \hline
\#11 & 4 & 150 & criterion\_track\_record, criterion\_relevance\_originality\_topicality, criterion\_feasibility, criterion\_suitability, candidate\_other, candidate\_quantity, proposal\_general, proposal\_method, positive, negative, suggestion, rationale, rationale\_context, impact\_beyond & Stratified sample of 30 sentences for each of the five evaluation criteria \\ \hline
\#12 & 4 & 1,000 & criterion\_track\_record, criterion\_relevance\_originality\_topicality, criterion\_feasibility, criterion\_suitability, candidate\_other, candidate\_quantity, proposal\_general, proposal\_method, positive, negative, suggestion, rationale, rationale\_context, impact\_beyond & Stratified sample of 200 sentences for each of the five evaluation criteria \\ \hline

\end{longtable}

Some categories from the annotation codebook have been renamed in the main text of the manuscript for easier interpretation. Table \ref{tab:mapping} below provides the mapping between the category names in the manuscript and the annotation codebook/implementation codes on GitHub.\\

\begin{table}[ht]
    \centering
    \begin{tabular}{lll}
    \toprule
    \textbf{Manuscript Category Names} & & \textbf{Codebook / GitHub Category Names} \\
    \midrule
        Track Record & & criterion\_track\_record \\
        Relevance, Originality, and Topicality & & criterion\_relevance\_originality\_topicality \\
        Suitability & & criterion\_suitability \\
        Feasibility & & criterion\_feasibility \\
        Applicant & & candidate\_other \\
        Applicant: Quantity & & candidate\_quantity \\
        Proposal & & proposal\_general \\
        Method & & proposal\_method \\
        Positive & & positive \\
        Negative & & negative \\
        Suggestion & & suggestion \\
        Rationale & & rationale \\
        \bottomrule
    \end{tabular}
    \caption{Mapping of Category Names}
    \label{tab:mapping}
\end{table}

\pagebreak

\section{Fine-tuning Settings}

\begin{table}[ht]
    \centering
    \begin{tabular}{lll}
    \toprule
    \textbf{Hyperparameter} & & \textbf{Value} \\
    \midrule
        Optimizer & & AdamW \\
        Learning rate & & 2e-5 \\
        Weight decay & & 0.01 \\
        Epochs & & 3 \\
        Batch size & & 10 \\
        \bottomrule
    \end{tabular}
    \caption{Fine-tuning settings}
    \label{tab:settings}
\end{table}

\section{Descriptive Statistics}\label{app:descriptives}

Below, we summarize the descriptive statistics for the annotated sample of 3,000 sentences.

\begin{figure}[h!]
\caption{The number of categories associated with each sentence}
\label{fig:distribution_appendix}
\centering
\includegraphics[width=1\textwidth]{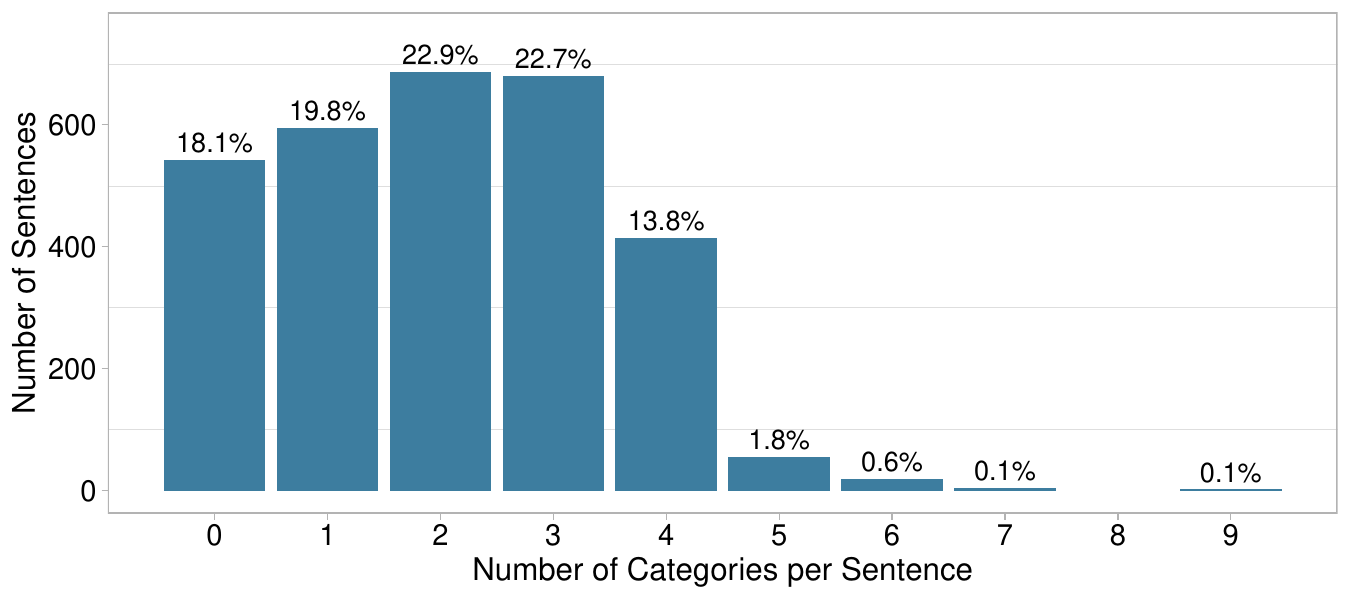}
\end{figure}

\begin{figure}[h!]
\caption{Prevalence of categories across the three coding rounds (each consisting of 1,000 sentences and each sentence being assessed by three of the four annotators), based on majority agreement}
\label{fig:class_prevalence_rounds}
\centering
\includegraphics[width=1\textwidth]{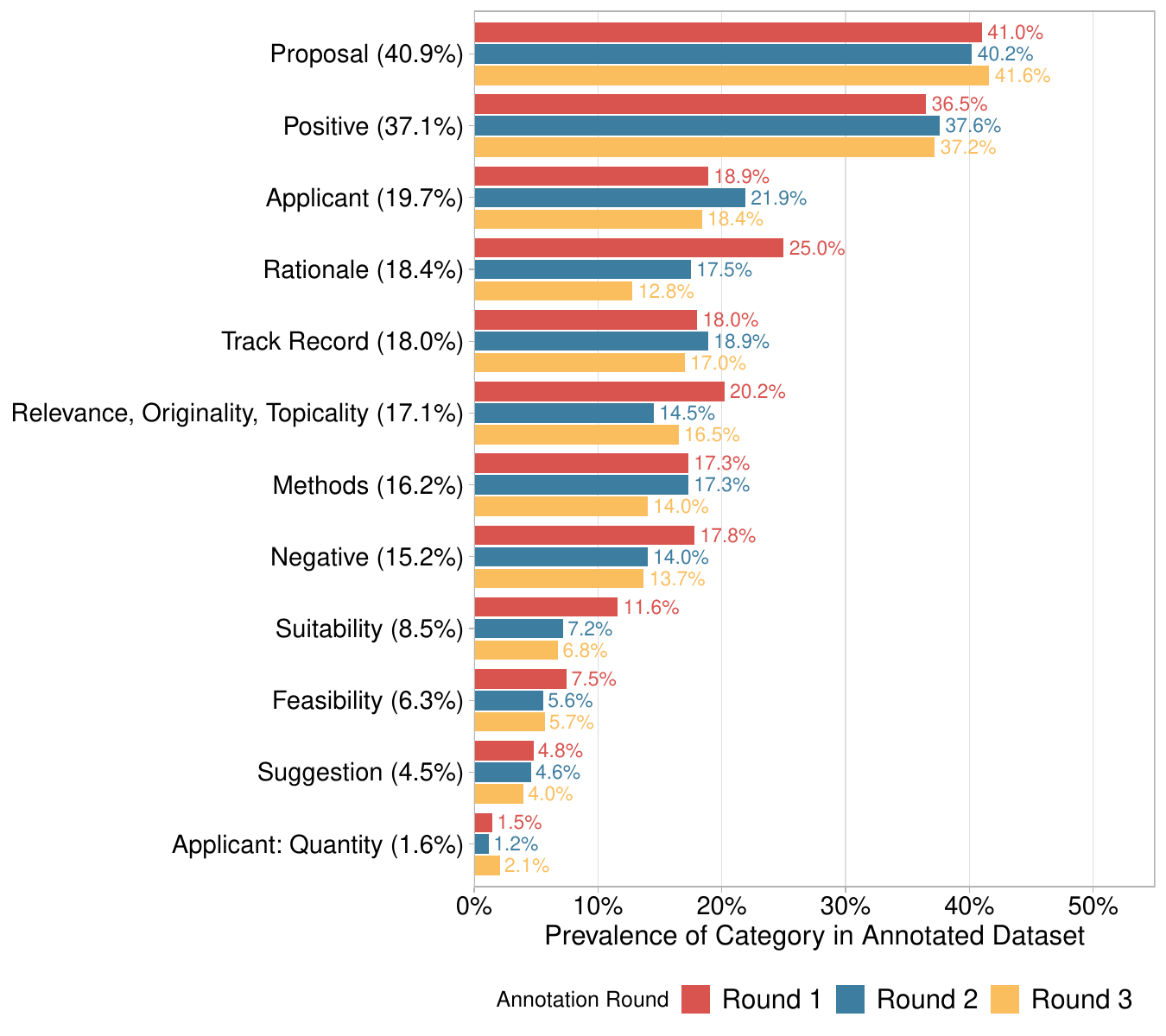}
\end{figure}

Each coding round consists of 1,000 sentences manually annotated by three coders. To determine whether the sentence is eventually considered under each category, we follow a ``majority agreement" process. In Figure \ref{fig:class_prevalence_rounds}, each bar corresponds to one coding round. Most of the categories maintain a similar prevalence across the three coding rounds with minor exceptions. 

The most apparent difference we find concerns the category \textit{Rationale}. This category is the most difficult category for the annotators as the existence of an argument or substantiation in a sentence is less clear to identify than other categories. In that sense, the prevalence for the category of rationale changed from 25\% in the first coding round to 12.8\% in the third round. It is necessary to consider that there are adjustments between annotators between rounds.

\pagebreak

\section{Robustness Tests and Additional Analyses} \label{sec:robustness}

\subsection{Detailed Overview of Classifier Performance} 

\label{sec:detailedoverview}

\begin{figure}[h!]
\caption{Macro F1 scores based on 5-fold cross-validation and single training-test split for binary classification. Cross-validation folds as well as the test set consist of 500 sentences. Grey dots show the minimum and maximum F1 Scores across the 5 folds. White circles show the average across the 5 folds.}
\label{fig:f1_compare}
\centering
\includegraphics[width=1\textwidth]{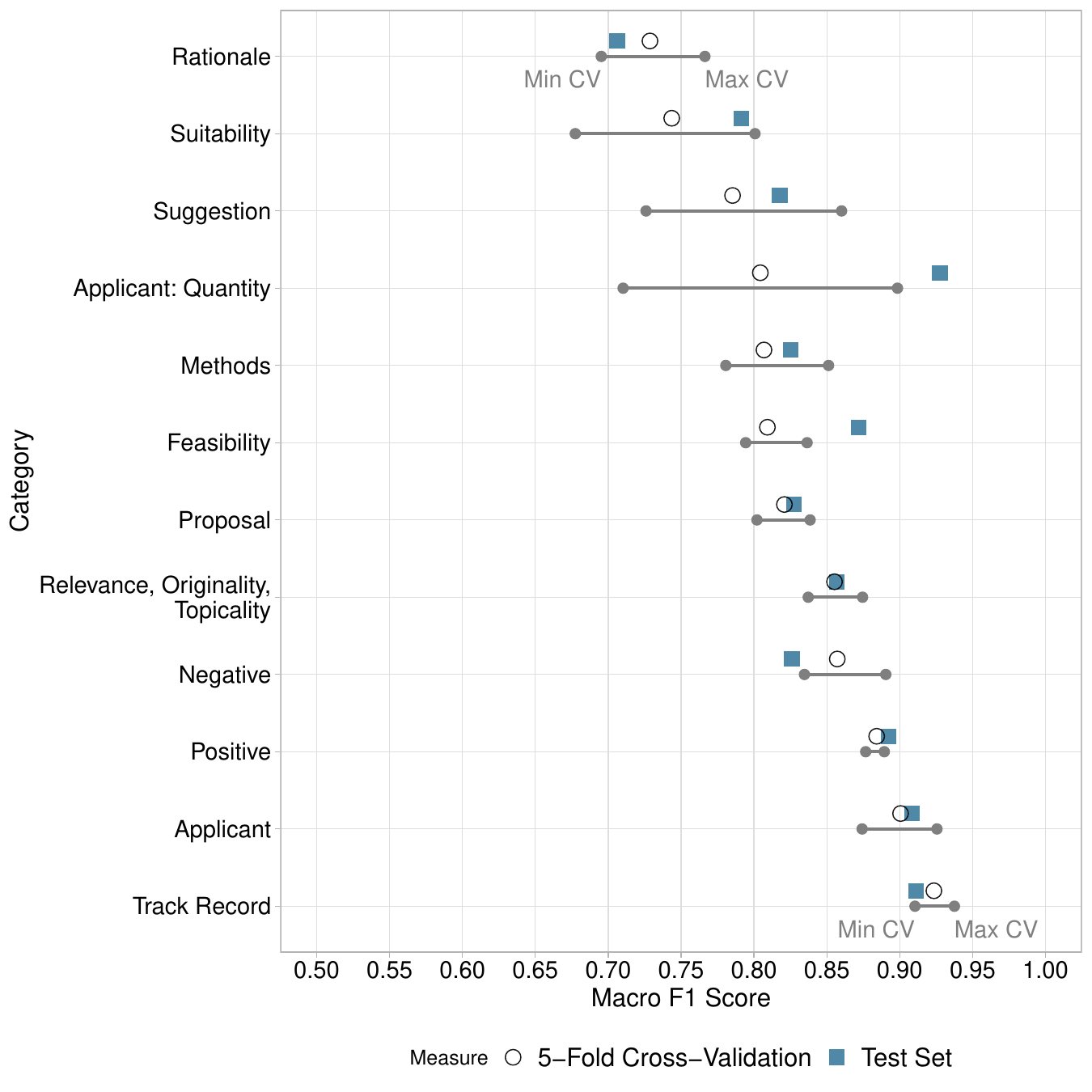}
\end{figure}

\begin{landscape}

\input{tables/table_f1_compare_detailed_binary.tex}

\input{tables/table_f1_compare_detailed_multilabel.tex}

\input{tables/table_f1_compare_detailed_multitask.tex}

 \input{tables/table_f1_compare_detailed_binary_sentences_with_full_agreement_only.tex}

\end{landscape}

\pagebreak

\subsection{Longer Context Evaluation}\label{app:context}

\begin{figure}[h!]
\caption{Prevalence of the rationale categories in an annotated set of 3,000 sentences: \textit{Rationale}, \textit{Rationale: Context}, \textit{Rationale and Context} and \textit{Rationale and Context} in \textit{Positive} and \textit{Negative} sentences. \textit{Note}: Dark blue colors show the percentages of sentences classified into the same category by all annotators; light blue colors indicate sentences classified as the same category by two of three annotators.}
\label{fig:rationale_context}
\centering
\includegraphics[width=1\textwidth]{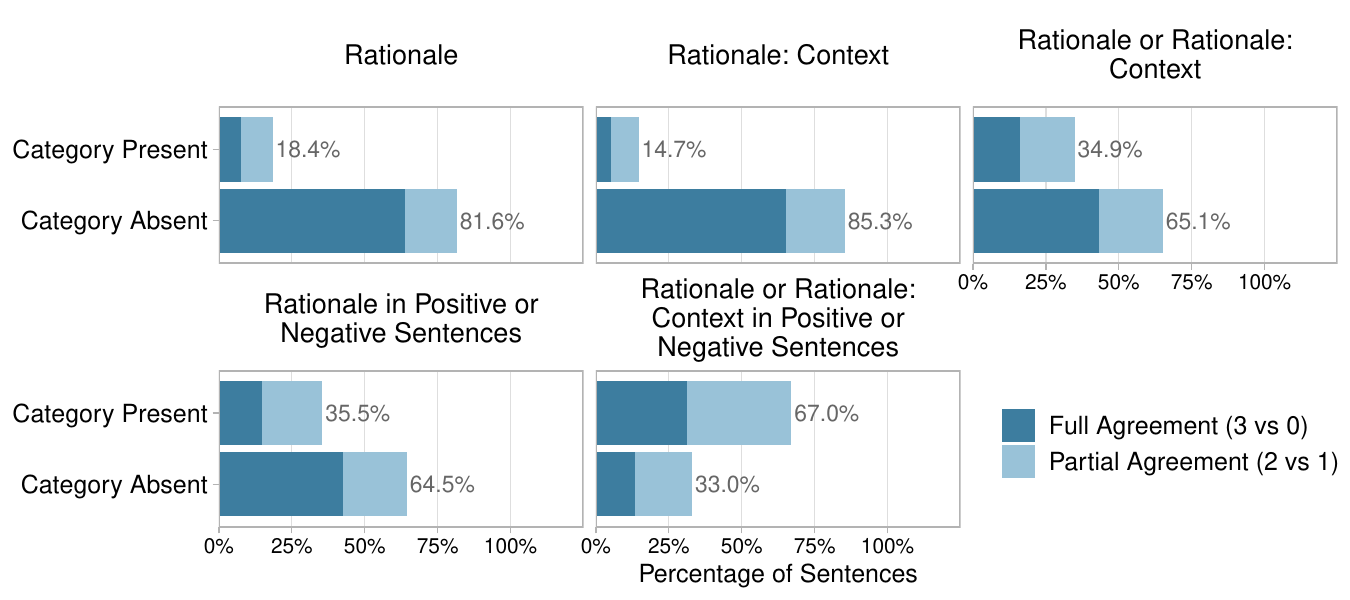}
\end{figure}

\input{tables/table_f1_compare_rationale.tex}

\begin{figure}[h!]
\caption{Macro F1 scores for prediction of \textit{Rationale} based on 5-fold cross-validations for different datasets. The dark blue color indicates results from a model fine-tuned only on one sentence (i.e., the main model used in the analysis). The light blue color indicates results from a classifier fine-tuned on \textit{Rationale and Context} with surrounding sentences. More details on the annotation are provided in the text. Circles show the minimum and maximum F1 scores; squares show the average F1 score.}
\label{fig:f1_compare_binary_rationale}
\centering
\includegraphics[width=1\textwidth]{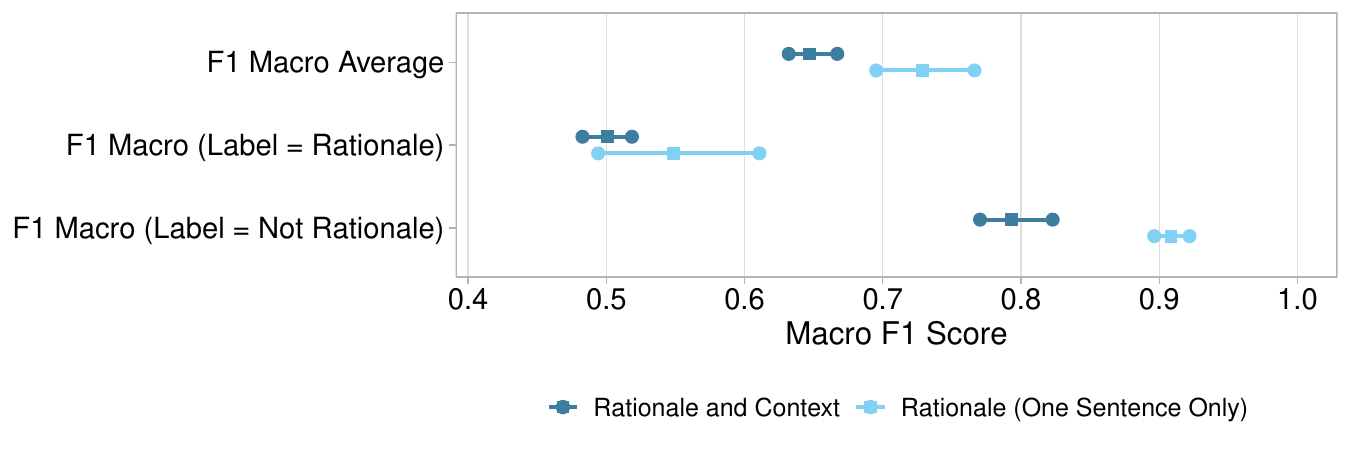}
\end{figure}

\clearpage

\subsection{Limiting Training Set to Sentences with Full Agreement}
\label{trainingvsfull}

\input{tables/table_f1_compare_binary_full.tex}

\begin{figure}[h!]
\caption{Macro F1 scores based on five-fold cross-validations for different training sets. Dark blue indicates the training set consisting only of sentences with full agreement. Light blue indicates the training set of all sentences. Circles show the minimum and maximum F1 scores; squares show the average F1 score.}
\label{fig:f1_compare_binary_full}
\centering
\includegraphics[width=1\textwidth]{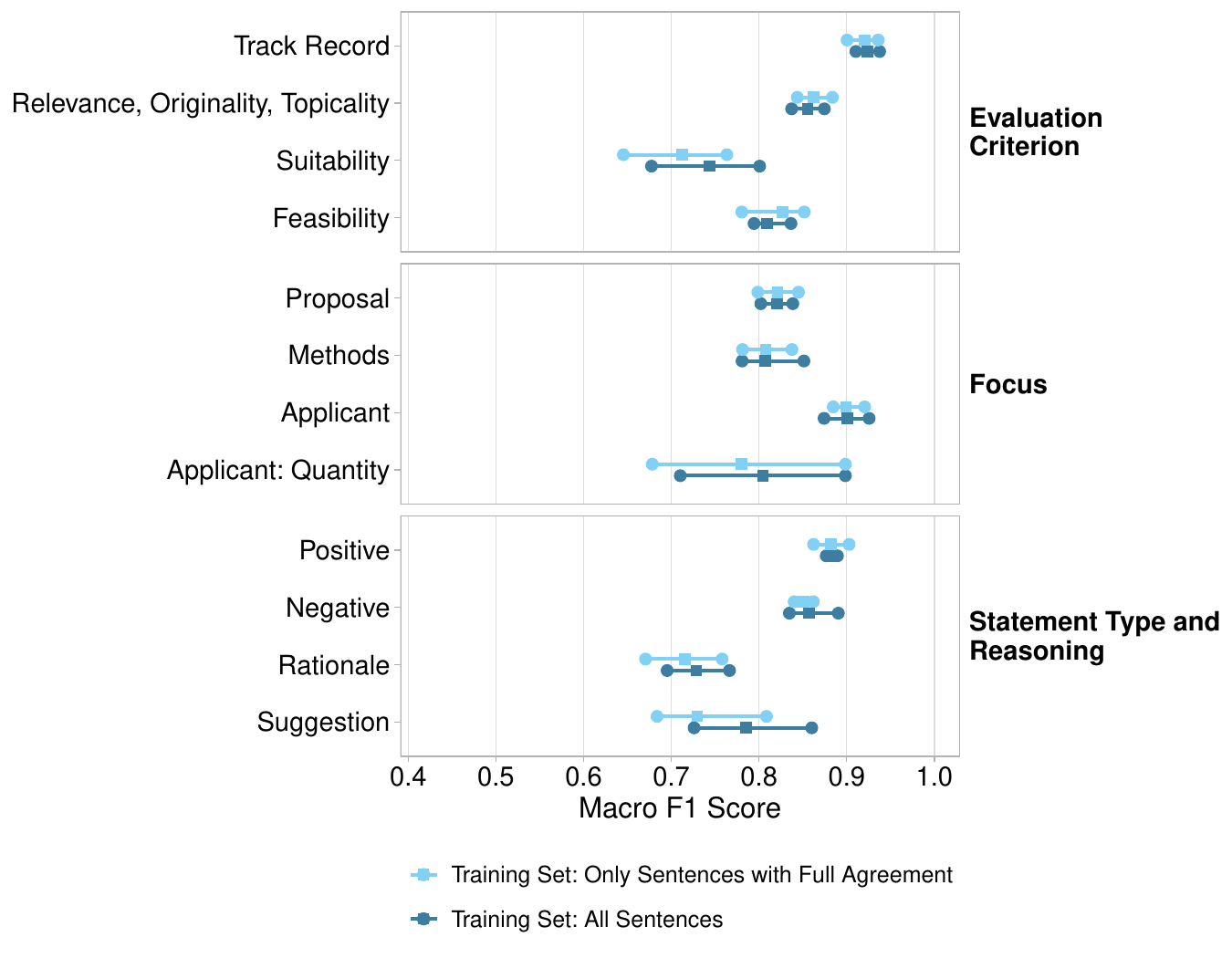}
\end{figure}

\clearpage

\subsection{Additional Training Data Via Few-Shot LLM Classification}\label{sec:llama}

Recently, Large Language Models (LLMs) have been increasingly employed to annotate additional data, which are then used to fine-tune smaller transformer models  \citep[see][]{Pelaez2024, Gupta2024}. We tested this approach via few-shot learning \citep{brown2020language} using the Llama model \citep{touvron2023llama}, in particular the \textsf{Meta-Llama-3-8B-Instruct} version \citep{llama3modelcard}, a Large Language Model that comes close to the performance of the best proprietary models. Yet, Llama 3 has the advantage that the smaller, more efficient version can be run locally, which is important to avoid data privacy concerns.

We tested the model's ability to classify the twelve categories by developing tailored prompts based on the Hugging Face chat template.\footnote{\url{https://huggingface.co/docs/transformers/main/chat_templating}.} Our prompts use the category descriptions based on the annotation codebook provided in Table \ref{tab:overview_categories}. As part of the prompt, we listed one example of a sentence representing given category according to all four human annotators, providing an unambiguous case for the category. We also listed one example of a sentence not representing any given category according to all four human annotators, providing a case with full agreement for the category. The complete prompt structure is available in Figure \ref{fig:chat_template} below. 

\begin{figure}[h!]
 \caption{Illustration of the prompt structure based on the Hugging Face chat template. The prompt is adjusted for each of the twelve categories. It provides the name of the category and its description as well as one full-agreement example sentence of the category and one full-agreement example sentence of no category. The last output from the 'Assistant', i.e. the $<$\textsf{GENERATED-LABEL}$>$ is the LLM prediction for the label of the given sentence from the test set.}
    \label{fig:chat_template}
\centering
\includegraphics[width=0.9\textwidth]{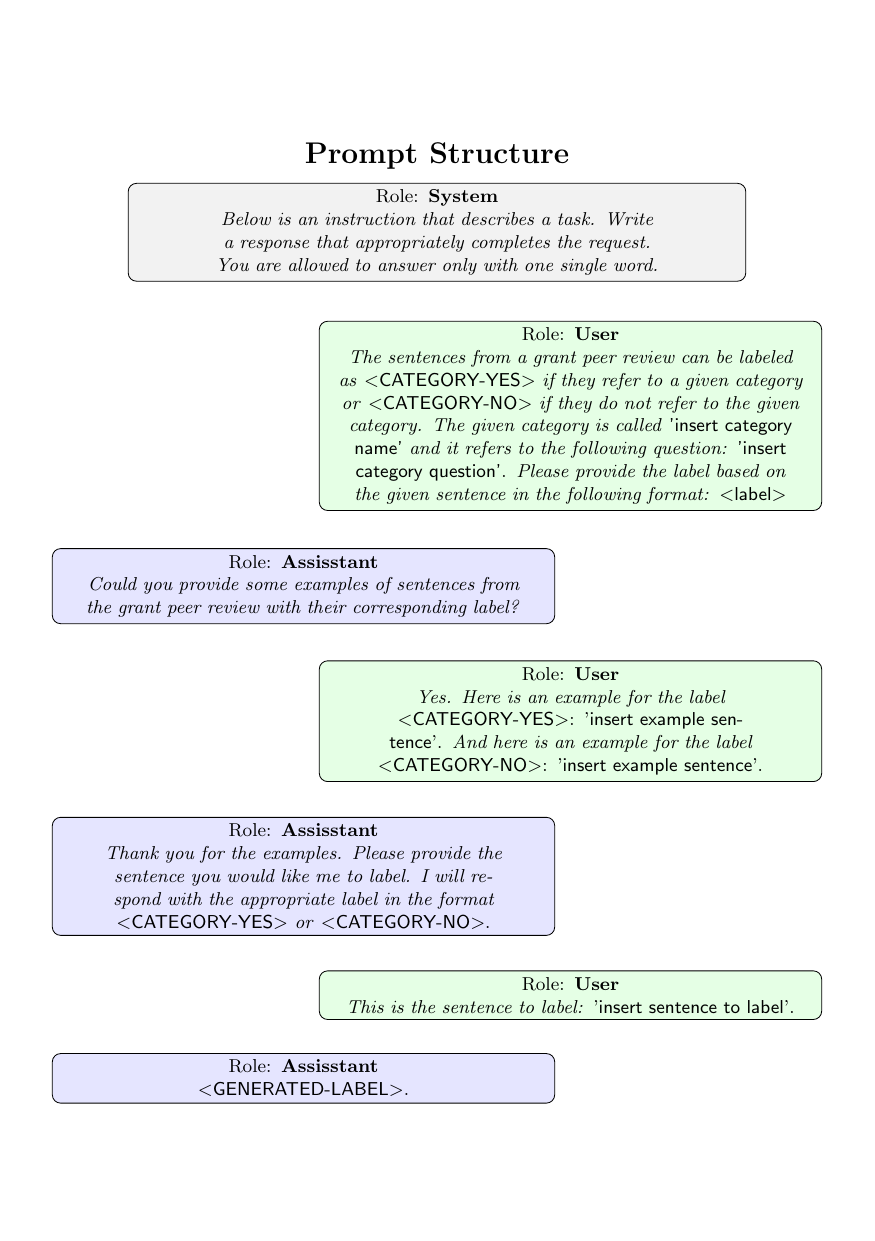}
\end{figure}

We used this prompt structure to classify the texts of the 500 sentences from the test set for a direct comparison with the results of the fine-tuned BERT models. Table \ref{tab:llama_annotation}  reveals that the classification performance of the Llama model is unanimously worse for all 12 categories. The average macro F1 score of 0.70 is substantially lower than the average score for the main SPECTER2 model (F1 score of 0.85). For a few categories, there is a rather small discrepancy, but for the majority of categories, there is an order-of-magnitude difference. The Llama model performed poorly, particularly in classifying sentences for imbalanced categories.

This further highlights that the annotation and classification of grant peer review sentences is a highly complex task. In this case, an LLM annotation does not match human annotation performance and neither does an LLM classification match the classification of fine-tuned BERT models. Recent studies by \citet{sun2023text} and \citet{bucher2024} provide similar findings, where fine-tuning BERT-based models on relatively small domain datasets outperforms the classification via classical few-shot learning of LLMs. For these reasons, we refrain from adding additional artificially generated training data to our models and from using LLMs for text classification directly.

\begin{landscape}
\input{tables/table_f1_compare_detailed_llama.tex}
\end{landscape}

\pagebreak

\section{Example Sentences for Each Category}
\label{sec:examplesentences}

Here, we show five example sentences for each category. These examples have been randomly selected from a sample of sentences where three coders agreed that the specific category is present.

\subsection{Criterion: Track Record}

\begin{itemize}
\item The applicants have considerable expertise in these approaches and present preliminary experimental data which support the proposal.
\item Moreover, she realized this impressing track record in a very competitive sub-discipline of REMOVE\_DISCIPLINE at relatively young age and in a period when she gave birth to and took care of three children.
\item The team in the [UNK] lab is outstanding with a long track record of meaningful publications elucidating the mechanisms of human and animal chlamydial infections.
\item She has the necessary (scarce) background in both REMOVE\_DISCIPLINE, has shown to be capable of doing interdisciplinary research and to go into themes which are novel.
\item This CV demonstrates the PIs track record of research excellence appropriate to carry out the project proposed.
\end{itemize}

\subsection{Criterion: Relevance, Originality, and Topicality}

\begin{itemize}
\item My judgment on the relevance and originality is similar as last time: I think the topic is highly relevant, original, and topical.
\item This is very the originality and novelty of the current proposal is situated.
\item The research proposed is relevant and pertinent, promising methodological advances of broad applicability by the community of REMOVE\_DISCIPLINE and scientists of REMOVE\_DISCIPLINE, but potentially to managers as well.
\item This project is very original and new for the PI.
\item In my opinion the main topic of the proposed project is scientifically very relevant.
\end{itemize}

\subsection{Criterion: Suitability}

\begin{itemize}
\item The meths are suitable and can be clearly achieved in this laboratory.
\item The suggested methodology reflects the state-of-the-art technologies with innovative approaches.
\item The project is well designed and the progress of preliminary work for the project shows the methodology and technology used in this project is robust and suitable.
\item The applicants have identified suitable ML techniques (CNN, Random Forest, etc).
\item Overall, I find the methodological approach absolutely convincing, and the applicant surely is competent to conduct the research as proposed.
\end{itemize}

\subsection{Criterion: Feasibility}

\begin{itemize}
\item The proposed project is feasible based on their expertise.
\item With eight clearly defined aims and methodologies, objectives are achievable.
\item The feasibility of the proposal is high.
\item Because I consider that the Applicant has addressed the criticism from the previous Reviewers in a satisfactory way and because of the Applicant’s previous work on the topic, I consider the work proposed feasible.
\item The proposed study is certainly feasible and has the potential to offer a very significant contribution to the academic community.
\end{itemize}

\subsection{Proposal}

\begin{itemize}
\item The proposal not only has a strong rationale, it also incorporates original ideas and questions that will be tested using modern approaches.
\item The proposal is highly compelling and the applicants have clearly presented the state-of-the-art in the field and how they will build on recent achievements in REMOVE\_DISCIPLINE.
\item The topic of study they propose is novel because, although parental effects have been studied since the 60s, the topic of REMOVE\_DISCIPLINE began in 2003 (current REMOVE\_DISCIPLINE) with invertebrates.
\item The project deals with an undoubtedly very interesting subject and its editorial parts are generally welcome.
\item The research described in the proposal is timely, novel, and has the potential to impact both research and practice.
\end{itemize}

\subsection{Methods}

\begin{itemize}
\item The reduced-form methods suggested in the proposal seem to be the appropriate method for the proposed projects.
\item The methodology of this project is very well articulated.
\item All methods proposed are either standard methods, already extensively established in the applicant's lab or provided by experienced collaborators.
\item In general, the applicant intends to use solid, well-developed data processing methodologies as a starting point, so that I have no doubts about their feasibility.
\item The applicant is using statistical and analytic approaches that are appropriate.
\end{itemize}

\subsection{Applicant}

\begin{itemize}
\item Therefore, they have ability to carry out the proposed project based on their strong expertise in this special research area and their outstanding track record.
\item The applicants [UNK] [UNK] and [UNK] [UNK] from REMOVE\_INSTITUTION (PSI) have done a lot of research on the non-destructive testing of ductile-to-brittle transition temperature of irradiated materials, theoretically analyzed the feasibility of the project, and provided some preliminary work results to support this project.
\item Their combined expertise has led them to propose a synthetic program of research that will have a much higher probability of novel contributions.
\item Clearly, the scientific research of Professor [UNK] [UNK] encompasses these important areas of REMOVE\_DISCIPLINE, REMOVE\_DISCIPLINE, and REMOVE\_DISCIPLINE, that a are relevant and necessary for this project.
\item The applicants are very well qualified for the proposed project.
\end{itemize}

\subsection{Applicant: Quantity}

\begin{itemize}
\item He is first author of 8 manuscripts and reviews and for four manuscript he is last author, and in addition he is joint last author on one additional manuscript.
\item Dr [UNK] has an h-index of 19, which is very good for a scientist this early in his/her career.
\item Dr [UNK] ’s research activity has yielded more than 420 papers in the most influential international journals and 40 patents, which made him ranked 53rd for the chemist list and 29th for the material scientists list in 2011.
\item The applicant has authored or co-authored 23 published articles in national and international scientific journals, and the work has been cited in 82 documents as of June 2018 on Scopus.
\item F. [UNK] has published nearly 80 articles in leading international journals over the period 2003 - 2017, which makes him a very productive researcher.
\end{itemize}

\subsection{Positive Statement}

\begin{itemize}
\item The strengths of this proposal are the strong applicant and team, the current dynamic nature of the field under study, and the potential impact of the work on advancing biological understanding and clinical care.
\item The proposal shows a good understanding of the state of the art in the area of REMOVE\_DISCIPLINE and quality transparency and the work is described accurately.
\item The applicant’s scientific record of accomplishment suggests that this project can be successfully performed and completed.
\item The applicant has made some outstanding contributions to the development of new methods in REMOVE\_DISCIPLINE, and the methods proposed here are beyond state of the art and fully appropriate for the tasks at hand.
\item This will be a very important project for future of the REMOVE\_DISCIPLINE design for the country.
\end{itemize}

\subsection{Negative Statement}

\begin{itemize}
\item There are a lot of spelling mistakes and references that are misleading or difficult to understand.
\item The PI appears to believe that NP = REMOVE\_DISCIPLINE cells, which is not the case.
\item The technique is not introduced and it appears as if the applicant wrote it only to someone from his own specific niche of REMOVE\_DISCIPLINE.
\item The lit review for the overall project seems somewhat inadequate, because it varies in the degree to which it is reasonably thorough for each of the substudies.
\item Thus it is questionable how we might expect these results to extend to the real world.
\end{itemize}

\subsection{Rationale}

\begin{itemize}
\item Given the comments above, this reviewer considers that the proposed methods are well suited to provide key answers to the proposed hypotheses, particularly because these methods and models have provided the background knowledge justifying the "reverse engineering" approach.
\item This is an interesting proposal, which combines a range of novel and well-established approaches to address an interesting question regarding the impacts of REMOVE\_DISCIPLINE on alpine streams via the mechanism of increased intermittency.
\item This proposal will have a broad impact upon REMOVE\_DISCIPLINE, REMOVE\_DISCIPLINE and REMOVE\_DISCIPLINE communities because the study of variability concerns to systematic and the result of the evolutionary studies of REMOVE\_DISCIPLINE is expect to show biogeographic patterns of interest to geoscientists.
\item The project has the potential to make an important contribution to knowledge, reporting behaviors that are positive or negative in relation to the Internet, and what could be the positive effect of the Internet for adolescent health.
\item Overall, the structure of the project is very detailed in relation to the proposed budget, demonstrates clearly the flow of information from data and experimental observations to modelling and then to engineering predictions, and clearly indicates the grasp of the applicant on the topic.
\end{itemize}

\subsection{Suggestion}

\begin{itemize}
\item If the project is accepted, then I would suggest that the project pays specific attention to increase practical relevance.
\item I feel a bit more details and clarification would be useful.
\item The research team may consider to do a sample size calculation and change the sample size accordingly.
\item It would be better to do standard addition (or matrix-matched standards) to eliminate any concerns.
\item I suggest to re-submit the proposal when the first steps of the two sub-projects are successfully done.
\end{itemize}

\pagebreak

\end{document}

%% file: tables/table_f1_compare_all_main.tex
\begin{table}
\centering
\caption{Comparison of F1-scores (macro-average) for binary, multi-label and multi-task classification. Note: Categories are sorted in
descending order of label shares. Colours highlight the order and value of F1-scores. The last row shows the average F1 scores across all categories.}\label{tab:f1_compare_all_main}
\centering
\fontsize{9}{11}\selectfont
\begin{tabular}[t]{l>{\centering\arraybackslash}p{1.5cm}>{\centering\arraybackslash}p{2cm}>{\centering\arraybackslash}p{2cm}>{\centering\arraybackslash}p{2cm}}
\toprule
\textbf{Category} & \textbf{Share Label} & \textbf{F1 \phantom{...}Binary\phantom{...}} & \textbf{F1 Multi-label} & \textbf{F1 Multi-task}\\
\midrule
Proposal & \textcolor{black}{40.9\%} & \cellcolor[HTML]{9ECCB7}{0.83} & \cellcolor[HTML]{A6D1BD}{0.80} & \cellcolor[HTML]{B5D9C8}{0.74}\\
\addlinespace
Positive & \textcolor{black}{37.1\%} & \cellcolor[HTML]{8CC2A9}{0.89} & \cellcolor[HTML]{9BCAB4}{0.84} & \cellcolor[HTML]{A2CEBA}{0.81}\\
\addlinespace
Applicant & \textcolor{black}{19.7\%} & \cellcolor[HTML]{88BFA6}{0.91} & \cellcolor[HTML]{8CC2A9}{0.90} & \cellcolor[HTML]{8BC1A8}{0.90}\\
\addlinespace
Rationale & \textcolor{black}{18.4\%} & \cellcolor[HTML]{BEDFD0}{0.71} & \cellcolor[HTML]{D2EADF}{0.63} & \cellcolor[HTML]{F5FFFA}{0.45}\\
\addlinespace
Track Record & \textcolor{black}{18.0\%} & \cellcolor[HTML]{88BFA6}{0.91} & \cellcolor[HTML]{89C0A7}{0.90} & \cellcolor[HTML]{88BFA6}{0.91}\\
\addlinespace
Relevance, Originality, Topicality & \textcolor{black}{17.1\%} & \cellcolor[HTML]{97C8B1}{0.86} & \cellcolor[HTML]{A3CFBB}{0.81} & \cellcolor[HTML]{D1EADE}{0.64}\\
\addlinespace
Methods & \textcolor{black}{16.2\%} & \cellcolor[HTML]{9FCDB8}{0.83} & \cellcolor[HTML]{B1D7C5}{0.76} & \cellcolor[HTML]{F5FFFA}{0.48}\\
\addlinespace
Negative & \textcolor{black}{15.2\%} & \cellcolor[HTML]{9ECCB7}{0.83} & \cellcolor[HTML]{96C7B0}{0.86} & \cellcolor[HTML]{D3EBE0}{0.63}\\
\addlinespace
Suitability & \textcolor{black}{8.5\%} & \cellcolor[HTML]{A7D2BE}{0.79} & \cellcolor[HTML]{C2E1D3}{0.69} & \cellcolor[HTML]{F5FFFA}{0.48}\\
\addlinespace
Feasibility & \textcolor{black}{6.3\%} & \cellcolor[HTML]{93C5AE}{0.87} & \cellcolor[HTML]{F5FFFA}{0.48} & \cellcolor[HTML]{F5FFFA}{0.48}\\
\addlinespace
Suggestion & \textcolor{black}{4.5\%} & \cellcolor[HTML]{A0CDB8}{0.82} & \cellcolor[HTML]{DDF1E8}{0.59} & \cellcolor[HTML]{F5FFFA}{0.49}\\
\addlinespace
Applicant: Quantity & \textcolor{black}{1.6\%} & \cellcolor[HTML]{83BCA2}{0.93} & \cellcolor[HTML]{F5FFFA}{0.50} & \cellcolor[HTML]{F5FFFA}{0.50}\\
\addlinespace
\midrule
Average F1 Score Across all Categories & \textcolor{black}{---} & \cellcolor[HTML]{99C9B3}{0.85} & \cellcolor[HTML]{B8DBCB}{0.73} & \cellcolor[HTML]{D4EBE0}{0.62}\\
\bottomrule
\end{tabular}
\end{table}

%% file: tables/table_f1_compare_binary_pretrained.tex
\begin{table}
\centering
\caption{\label{tab:tab:f1_compare_binary_pretrained}Comparison of F1-scores (macro-average) for binary classifiers pre-trained on SPECTER2, BERT, and RoBERTa. Note: Categories are sorted in
descending order of label shares. Colours highlight the order and value of F1-scores. The last row shows the average F1 scores across all categories.}
\centering
\fontsize{9}{11}\selectfont
\begin{tabular}[t]{l>{\centering\arraybackslash}p{1.5cm}>{\centering\arraybackslash}p{2cm}>{\centering\arraybackslash}p{2cm}>{\centering\arraybackslash}p{2cm}}
\toprule
\textbf{Category} & \textbf{Share Label} & \textbf{F1: SPECTER2} & \textbf{F1: \phantom{...}BERT\phantom{...}} & \textbf{F1: RoBERTa}\\
\midrule
Proposal & \textcolor{black}{40.9\%} & \cellcolor[HTML]{9ECCB7}{0.83} & \cellcolor[HTML]{9ACAB3}{0.84} & \cellcolor[HTML]{9ACAB3}{0.84}\\
\addlinespace
Positive & \textcolor{black}{37.1\%} & \cellcolor[HTML]{8CC2A9}{0.89} & \cellcolor[HTML]{88BFA6}{0.91} & \cellcolor[HTML]{89C0A7}{0.91}\\
\addlinespace
Applicant & \textcolor{black}{19.7\%} & \cellcolor[HTML]{88BFA6}{0.91} & \cellcolor[HTML]{8BC1A8}{0.90} & \cellcolor[HTML]{8BC1A8}{0.90}\\
\addlinespace
Rationale & \textcolor{black}{18.4\%} & \cellcolor[HTML]{BEDFD0}{0.71} & \cellcolor[HTML]{BEDFD0}{0.71} & \cellcolor[HTML]{BADCCC}{0.72}\\
\addlinespace
Track Record & \textcolor{black}{18.0\%} & \cellcolor[HTML]{88BFA6}{0.91} & \cellcolor[HTML]{83BCA2}{0.93} & \cellcolor[HTML]{84BDA3}{0.92}\\
\addlinespace
Relevance, Originality, Topicality & \textcolor{black}{17.1\%} & \cellcolor[HTML]{97C8B1}{0.86} & \cellcolor[HTML]{97C8B1}{0.86} & \cellcolor[HTML]{96C7B0}{0.86}\\
\addlinespace
Methods & \textcolor{black}{16.2\%} & \cellcolor[HTML]{9FCDB8}{0.83} & \cellcolor[HTML]{A0CDB8}{0.82} & \cellcolor[HTML]{98C8B2}{0.85}\\
\addlinespace
Negative & \textcolor{black}{15.2\%} & \cellcolor[HTML]{9ECCB7}{0.83} & \cellcolor[HTML]{95C7B0}{0.86} & \cellcolor[HTML]{98C8B2}{0.85}\\
\addlinespace
Suitability & \textcolor{black}{8.5\%} & \cellcolor[HTML]{A7D2BE}{0.79} & \cellcolor[HTML]{B1D7C5}{0.76} & \cellcolor[HTML]{A7D2BE}{0.79}\\
\addlinespace
Feasibility & \textcolor{black}{6.3\%} & \cellcolor[HTML]{93C5AE}{0.87} & \cellcolor[HTML]{9DCBB6}{0.83} & \cellcolor[HTML]{92C5AD}{0.87}\\
\addlinespace
Suggestion & \textcolor{black}{4.5\%} & \cellcolor[HTML]{A0CDB8}{0.82} & \cellcolor[HTML]{99C9B3}{0.85} & \cellcolor[HTML]{94C6AF}{0.87}\\
\addlinespace
Applicant: Quantity & \textcolor{black}{1.6\%} & \cellcolor[HTML]{83BCA2}{0.93} & \cellcolor[HTML]{81BBA0}{0.94} & \cellcolor[HTML]{9DCBB6}{0.83}\\
\addlinespace
\midrule
Average F1 Score Across all Categories & \textcolor{black}{---} & \cellcolor[HTML]{99C9B3}{0.85} & \cellcolor[HTML]{98C8B2}{0.85} & \cellcolor[HTML]{98C8B2}{0.85}\\
\bottomrule
\end{tabular}
\end{table}

%% file: tables/keyness_fullsample.tex
\begin{table}[!h]
\centering
\caption{Keyness analysis revealing predictive terms for each category in the full sample of project funding review reports (’unk’ is a placeholder for the applicant, co-applicant and project partner names to maintain anonymity)} 
\label{tab:keyness}
\begingroup\footnotesize
\begin{tabular}{p{0.35\textwidth}p{0.65\textwidth}}
  \hline
Category & Terms \\ 
  \hline \cellcolor{snsfblue!50} 
Track Record & unk, expertise, applicant, dr, publications, experience, track\_record, field, pi, published, prof, carry, team, journals, papers, expert, professor, university, excellent\_track\_record, research, phd, publication\_record, strong\_track\_record, excellent, applicants \\ 
   \hdashline  \cellcolor{snsfblue!50} 
Relevance, Originality, Topicality & originality, original, novel, timely, topicality, topical, scientific\_relevance, topic, relevance, relevant, innovative, highly\_relevant, understanding, novelty, potential, interesting, discipline, field, important, scientifically\_relevant, exciting, proposed\_project, project, interest, highly\_original \\ 
   \hdashline \cellcolor{snsfblue!50} 
Suitability & methods, suitable, appropriate, suitability, proposed\_methods, methodology, methods\_proposed, well\_suited, suited, method, answering, chosen\_methods, questions\_set, methods\_chosen, adequate, chosen, choice, approach, feasible, proposed\_methodology, sample\_size, well-suited, answer, use, methodological\_approach \\ 
   \hdashline \cellcolor{snsfblue!50} 
Feasibility & feasible, feasibility, milestones, timeline, workload, project, budget, personnel, planned\_duration, schedule, proportionate, given\_time, achievable, reasonable, time, realistic, available\_resources, timeframe, time\_frame, planned, ambitious, milestones\_set, reached, highly\_feasible, resources \\ 
   \hdashline  \cellcolor{snsfgreen!50}
Proposal & project, proposal, originality, topic, topicality, scientific\_relevance, timely, topical, original, proposed\_project, relevance, potential, ambitious, overall, aims, impact, scope, interesting, milestones, research\_plan, application, relevant, project\_aims, highly\_relevant, discipline \\ 
   \hdashline \cellcolor{snsfgreen!50}
Methods & methods, method, use, used, methodology, approach, suitability, suitable, proposed\_methods, using, experiments, model, samples, methods\_proposed, appropriate, data, sampling, sample\_size, models, approaches, sample, measurements, test, chosen\_methods, combination \\ 
   \hdashline \cellcolor{snsfgreen!50}
Applicant & unk, expertise, applicant, dr, experience, pi, team, prof, track\_record, field, carry, expert, applicants, professor, university, collaborators, publications, research, excellent\_track\_record, excellent, strong\_track\_record, research\_team, phd, group, leader \\ 
   \hdashline \cellcolor{snsfgreen!50}
Applicant: Quantity & published, h-index, publications, papers, citations, first\_author, articles, journals, last\_five\_years, last\_author, last\_years, google\_scholar, senior\_author, scopus, corresponding\_author, published\_papers, co-author, co-authored, pubmed, authored, published\_articles, past\_years, journal, peer-reviewed\_journals, book\_chapters \\ 
   \hdashline  \cellcolor{snsfred!50}
Positive & field, project, expertise, excellent, unk, original, applicant, overall, proposed\_project, relevant, timely, good, carry, outstanding, highly\_relevant, strong, research, team, feasible, topical, interesting, originality, topic, track\_record, innovative \\ 
   \hdashline \cellcolor{snsfred!50}
Negative & lack, however, unclear, clear, difficult, little, missing, weakness, limited, concern, seems, somewhat, bit, seem, unfortunately, lacks, vague, lacking, proposal, although, details, weaknesses, description, concerns, mention \\ 
   \hdashline \cellcolor{snsfred!50}
Rationale & lack, due, including, given, etc, fact, evidenced, since, different, numerous, description, particularly, providing, nature, historical, processes, understanding, history, clear, publications, demonstrating, data, making, absence, state-of-the-art \\ 
   \hdashline \cellcolor{snsfred!50}
Suggestion & helpful, better, suggest, recommend, strengthened, useful, consider, strengthen, clearer, encourage, might, stronger, beneficial, perhaps, advisable, benefit, think, include, benefited, desirable, detail, recommended, preferable, advise, suggestion \\ 
   \hline
\end{tabular}
\endgroup
\end{table}

%% file: tables/table_f1_compare_detailed_binary.tex
\begin{table}
\centering
\caption{Detailed overview of the classification test results based on binary classification. Note:  Categories are sorted in descending order of label shares.}\label{tab:f1_compare_detailed_binary}
\centering
\fontsize{6}{8}\selectfont
\begin{tabular}[t]{l>{\centering\arraybackslash}p{1cm}>{\centering\arraybackslash}p{1cm}>{\centering\arraybackslash}p{1cm}>{\centering\arraybackslash}p{1cm}>{\centering\arraybackslash}p{1cm}>{\centering\arraybackslash}p{1cm}>{\centering\arraybackslash}p{1cm}>{\centering\arraybackslash}p{1cm}>{\centering\arraybackslash}p{1cm}>{\centering\arraybackslash}p{1cm}>{\centering\arraybackslash}p{1cm}>{\centering\arraybackslash}p{1cm}>{\centering\arraybackslash}p{1cm}>{\centering\arraybackslash}p{1cm}>{\centering\arraybackslash}p{1cm}}
\toprule
\textbf{Category} & \textbf{Share Label} & \textbf{Acc.} & \textbf{Bal. Acc.} & \textbf{F1 (Macro)} & \textbf{F1 (Micro)} & \textbf{F1 Lab=1} & \textbf{F1 Lab=0} & \textbf{Prec. (Macro)} & \textbf{Prec. (Micro)} & \textbf{Prec. Lab=1} & \textbf{Prec. Lab=0} & \textbf{Recall (Macro)} & \textbf{Recall (Micro)} & \textbf{Recall Lab=1} & \textbf{Recall Lab=0}\\
\midrule
Proposal & \textcolor{black}{40.9\%} & \textcolor{black}{0.83} & \textcolor{black}{0.83} & \cellcolor[HTML]{9ECCB7}{0.83} & \textcolor{black}{0.83} & \cellcolor[HTML]{A7D2BE}{0.79} & \cellcolor[HTML]{95C7B0}{0.86} & \textcolor{black}{0.83} & \textcolor{black}{0.83} & \textcolor{black}{0.81} & \textcolor{black}{0.85} & \textcolor{black}{0.83} & \textcolor{black}{0.83} & \textcolor{black}{0.78} & \textcolor{black}{0.87}\\
\addlinespace
Positive & \textcolor{black}{37.1\%} & \textcolor{black}{0.90} & \textcolor{black}{0.90} & \cellcolor[HTML]{8CC2A9}{0.89} & \textcolor{black}{0.90} & \cellcolor[HTML]{94C6AF}{0.87} & \cellcolor[HTML]{86BEA4}{0.92} & \textcolor{black}{0.89} & \textcolor{black}{0.90} & \textcolor{black}{0.84} & \textcolor{black}{0.94} & \textcolor{black}{0.90} & \textcolor{black}{0.90} & \textcolor{black}{0.90} & \textcolor{black}{0.90}\\
\addlinespace
Applicant & \textcolor{black}{19.7\%} & \textcolor{black}{0.94} & \textcolor{black}{0.92} & \cellcolor[HTML]{88BFA6}{0.91} & \textcolor{black}{0.94} & \cellcolor[HTML]{97C8B1}{0.85} & \cellcolor[HTML]{7BB89C}{0.96} & \textcolor{black}{0.90} & \textcolor{black}{0.94} & \textcolor{black}{0.82} & \textcolor{black}{0.97} & \textcolor{black}{0.92} & \textcolor{black}{0.94} & \textcolor{black}{0.89} & \textcolor{black}{0.95}\\
\addlinespace
Rationale & \textcolor{black}{18.4\%} & \textcolor{black}{0.83} & \textcolor{black}{0.70} & \cellcolor[HTML]{BEDFD0}{0.71} & \textcolor{black}{0.83} & \cellcolor[HTML]{F0FCF6}{0.52} & \cellcolor[HTML]{8CC2A9}{0.90} & \textcolor{black}{0.71} & \textcolor{black}{0.83} & \textcolor{black}{0.53} & \textcolor{black}{0.89} & \textcolor{black}{0.70} & \textcolor{black}{0.83} & \textcolor{black}{0.50} & \textcolor{black}{0.90}\\
\addlinespace
Track Record & \textcolor{black}{18.0\%} & \textcolor{black}{0.95} & \textcolor{black}{0.92} & \cellcolor[HTML]{88BFA6}{0.91} & \textcolor{black}{0.95} & \cellcolor[HTML]{97C8B1}{0.86} & \cellcolor[HTML]{79B69A}{0.97} & \textcolor{black}{0.90} & \textcolor{black}{0.95} & \textcolor{black}{0.82} & \textcolor{black}{0.98} & \textcolor{black}{0.92} & \textcolor{black}{0.95} & \textcolor{black}{0.89} & \textcolor{black}{0.96}\\
\addlinespace
Relevance, Originality, Topicality & \textcolor{black}{17.1\%} & \textcolor{black}{0.92} & \textcolor{black}{0.86} & \cellcolor[HTML]{97C8B1}{0.86} & \textcolor{black}{0.92} & \cellcolor[HTML]{AFD6C4}{0.76} & \cellcolor[HTML]{7EB99E}{0.95} & \textcolor{black}{0.85} & \textcolor{black}{0.92} & \textcolor{black}{0.75} & \textcolor{black}{0.95} & \textcolor{black}{0.86} & \textcolor{black}{0.92} & \textcolor{black}{0.78} & \textcolor{black}{0.95}\\
\addlinespace
Methods & \textcolor{black}{16.2\%} & \textcolor{black}{0.91} & \textcolor{black}{0.82} & \cellcolor[HTML]{9FCDB8}{0.83} & \textcolor{black}{0.91} & \cellcolor[HTML]{BEDFD0}{0.71} & \cellcolor[HTML]{7FBA9F}{0.95} & \textcolor{black}{0.84} & \textcolor{black}{0.91} & \textcolor{black}{0.73} & \textcolor{black}{0.94} & \textcolor{black}{0.82} & \textcolor{black}{0.91} & \textcolor{black}{0.68} & \textcolor{black}{0.95}\\
\addlinespace
Negative & \textcolor{black}{15.2\%} & \textcolor{black}{0.91} & \textcolor{black}{0.84} & \cellcolor[HTML]{9ECCB7}{0.83} & \textcolor{black}{0.91} & \cellcolor[HTML]{BDDECF}{0.71} & \cellcolor[HTML]{7FBA9F}{0.94} & \textcolor{black}{0.81} & \textcolor{black}{0.91} & \textcolor{black}{0.67} & \textcolor{black}{0.95} & \textcolor{black}{0.84} & \textcolor{black}{0.91} & \textcolor{black}{0.75} & \textcolor{black}{0.93}\\
\addlinespace
Suitability & \textcolor{black}{8.5\%} & \textcolor{black}{0.94} & \textcolor{black}{0.77} & \cellcolor[HTML]{A7D2BE}{0.79} & \textcolor{black}{0.94} & \cellcolor[HTML]{D6EDE2}{0.62} & \cellcolor[HTML]{79B69A}{0.97} & \textcolor{black}{0.82} & \textcolor{black}{0.94} & \textcolor{black}{0.69} & \textcolor{black}{0.96} & \textcolor{black}{0.77} & \textcolor{black}{0.94} & \textcolor{black}{0.56} & \textcolor{black}{0.98}\\
\addlinespace
Feasibility & \textcolor{black}{6.3\%} & \textcolor{black}{0.97} & \textcolor{black}{0.85} & \cellcolor[HTML]{93C5AE}{0.87} & \textcolor{black}{0.97} & \cellcolor[HTML]{B1D7C5}{0.76} & \cellcolor[HTML]{75B497}{0.99} & \textcolor{black}{0.90} & \textcolor{black}{0.97} & \textcolor{black}{0.81} & \textcolor{black}{0.98} & \textcolor{black}{0.85} & \textcolor{black}{0.97} & \textcolor{black}{0.71} & \textcolor{black}{0.99}\\
\addlinespace
Suggestion & \textcolor{black}{4.5\%} & \textcolor{black}{0.97} & \textcolor{black}{0.79} & \cellcolor[HTML]{A0CDB8}{0.82} & \textcolor{black}{0.97} & \cellcolor[HTML]{CDE7DB}{0.65} & \cellcolor[HTML]{75B497}{0.99} & \textcolor{black}{0.85} & \textcolor{black}{0.97} & \textcolor{black}{0.72} & \textcolor{black}{0.98} & \textcolor{black}{0.79} & \textcolor{black}{0.97} & \textcolor{black}{0.59} & \textcolor{black}{0.99}\\
\addlinespace
Applicant: Quantity & \textcolor{black}{1.6\%} & \textcolor{black}{1.00} & \textcolor{black}{0.88} & \cellcolor[HTML]{83BCA2}{0.93} & \textcolor{black}{1.00} & \cellcolor[HTML]{96C7B0}{0.86} & \cellcolor[HTML]{72B294}{1.00} & \textcolor{black}{1.00} & \textcolor{black}{1.00} & \textcolor{black}{1.00} & \textcolor{black}{1.00} & \textcolor{black}{0.88} & \textcolor{black}{1.00} & \textcolor{black}{0.75} & \textcolor{black}{1.00}\\
\bottomrule
\end{tabular}
\end{table}

%% file: tables/table_f1_compare_detailed_multilabel.tex
\begin{table}
\centering
\caption{\label{tab:tab:f1_compare_detailed_multilabel}Detailed overview of the classification test results based on multi-label classification. Note:  Categories are sorted in descending order of label shares.}
\centering
\fontsize{6}{8}\selectfont
\begin{tabular}[t]{l>{\centering\arraybackslash}p{1cm}>{\centering\arraybackslash}p{1cm}>{\centering\arraybackslash}p{1cm}>{\centering\arraybackslash}p{1cm}>{\centering\arraybackslash}p{1cm}>{\centering\arraybackslash}p{1cm}>{\centering\arraybackslash}p{1cm}>{\centering\arraybackslash}p{1cm}>{\centering\arraybackslash}p{1cm}>{\centering\arraybackslash}p{1cm}>{\centering\arraybackslash}p{1cm}>{\centering\arraybackslash}p{1cm}>{\centering\arraybackslash}p{1cm}>{\centering\arraybackslash}p{1cm}>{\centering\arraybackslash}p{1cm}}
\toprule
\textbf{Category} & \textbf{Share Label} & \textbf{Acc.} & \textbf{Bal. Acc.} & \textbf{F1 (Macro)} & \textbf{F1 (Micro)} & \textbf{F1 Lab=1} & \textbf{F1 Lab=0} & \textbf{Prec. (Macro)} & \textbf{Prec. (Micro)} & \textbf{Prec. Lab=1} & \textbf{Prec. Lab=0} & \textbf{Recall (Macro)} & \textbf{Recall (Micro)} & \textbf{Recall Lab=1} & \textbf{Recall Lab=0}\\
\midrule
Proposal & \textcolor{black}{40.9\%} & \textcolor{black}{0.81} & \textcolor{black}{0.79} & \cellcolor[HTML]{A6D1BD}{0.80} & \textcolor{black}{0.81} & \cellcolor[HTML]{B2D8C6}{0.75} & \cellcolor[HTML]{9ACAB3}{0.84} & \textcolor{black}{0.81} & \textcolor{black}{0.81} & \textcolor{black}{0.83} & \textcolor{black}{0.79} & \textcolor{black}{0.79} & \textcolor{black}{0.81} & \textcolor{black}{0.68} & \textcolor{black}{0.90}\\
\addlinespace
Positive & \textcolor{black}{37.1\%} & \textcolor{black}{0.85} & \textcolor{black}{0.84} & \cellcolor[HTML]{9BCAB4}{0.84} & \textcolor{black}{0.85} & \cellcolor[HTML]{A4D0BC}{0.80} & \cellcolor[HTML]{92C5AD}{0.87} & \textcolor{black}{0.84} & \textcolor{black}{0.85} & \textcolor{black}{0.82} & \textcolor{black}{0.86} & \textcolor{black}{0.84} & \textcolor{black}{0.85} & \textcolor{black}{0.79} & \textcolor{black}{0.88}\\
\addlinespace
Applicant & \textcolor{black}{19.7\%} & \textcolor{black}{0.93} & \textcolor{black}{0.90} & \cellcolor[HTML]{8CC2A9}{0.90} & \textcolor{black}{0.93} & \cellcolor[HTML]{9CCBB5}{0.83} & \cellcolor[HTML]{7CB89C}{0.96} & \textcolor{black}{0.89} & \textcolor{black}{0.93} & \textcolor{black}{0.82} & \textcolor{black}{0.96} & \textcolor{black}{0.90} & \textcolor{black}{0.93} & \textcolor{black}{0.85} & \textcolor{black}{0.95}\\
\addlinespace
Rationale & \textcolor{black}{18.4\%} & \textcolor{black}{0.84} & \textcolor{black}{0.61} & \cellcolor[HTML]{D2EADF}{0.63} & \textcolor{black}{0.84} & \cellcolor[HTML]{F7FFFB}{0.35} & \cellcolor[HTML]{88BFA6}{0.91} & \textcolor{black}{0.74} & \textcolor{black}{0.84} & \textcolor{black}{0.63} & \textcolor{black}{0.86} & \textcolor{black}{0.61} & \textcolor{black}{0.84} & \textcolor{black}{0.25} & \textcolor{black}{0.97}\\
\addlinespace
Track Record & \textcolor{black}{18.0\%} & \textcolor{black}{0.94} & \textcolor{black}{0.94} & \cellcolor[HTML]{89C0A7}{0.90} & \textcolor{black}{0.94} & \cellcolor[HTML]{9ACAB3}{0.84} & \cellcolor[HTML]{7AB79B}{0.96} & \textcolor{black}{0.88} & \textcolor{black}{0.94} & \textcolor{black}{0.77} & \textcolor{black}{0.98} & \textcolor{black}{0.94} & \textcolor{black}{0.94} & \textcolor{black}{0.93} & \textcolor{black}{0.94}\\
\addlinespace
Relevance, Originality, Topicality & \textcolor{black}{17.1\%} & \textcolor{black}{0.90} & \textcolor{black}{0.77} & \cellcolor[HTML]{A3CFBB}{0.81} & \textcolor{black}{0.90} & \cellcolor[HTML]{C6E4D6}{0.68} & \cellcolor[HTML]{80BBA0}{0.94} & \textcolor{black}{0.87} & \textcolor{black}{0.90} & \textcolor{black}{0.84} & \textcolor{black}{0.91} & \textcolor{black}{0.77} & \textcolor{black}{0.90} & \textcolor{black}{0.57} & \textcolor{black}{0.98}\\
\addlinespace
Methods & \textcolor{black}{16.2\%} & \textcolor{black}{0.88} & \textcolor{black}{0.72} & \cellcolor[HTML]{B1D7C5}{0.76} & \textcolor{black}{0.88} & \cellcolor[HTML]{DFF2E9}{0.58} & \cellcolor[HTML]{82BCA1}{0.93} & \textcolor{black}{0.82} & \textcolor{black}{0.88} & \textcolor{black}{0.74} & \textcolor{black}{0.90} & \textcolor{black}{0.72} & \textcolor{black}{0.88} & \textcolor{black}{0.48} & \textcolor{black}{0.97}\\
\addlinespace
Negative & \textcolor{black}{15.2\%} & \textcolor{black}{0.94} & \textcolor{black}{0.82} & \cellcolor[HTML]{96C7B0}{0.86} & \textcolor{black}{0.94} & \cellcolor[HTML]{B2D8C6}{0.75} & \cellcolor[HTML]{7AB79B}{0.96} & \textcolor{black}{0.92} & \textcolor{black}{0.94} & \textcolor{black}{0.89} & \textcolor{black}{0.94} & \textcolor{black}{0.82} & \textcolor{black}{0.94} & \textcolor{black}{0.65} & \textcolor{black}{0.99}\\
\addlinespace
Suitability & \textcolor{black}{8.5\%} & \textcolor{black}{0.92} & \textcolor{black}{0.64} & \cellcolor[HTML]{C2E1D3}{0.69} & \textcolor{black}{0.92} & \cellcolor[HTML]{F6FFFA}{0.42} & \cellcolor[HTML]{7CB89C}{0.96} & \textcolor{black}{0.86} & \textcolor{black}{0.92} & \textcolor{black}{0.79} & \textcolor{black}{0.92} & \textcolor{black}{0.64} & \textcolor{black}{0.92} & \textcolor{black}{0.29} & \textcolor{black}{0.99}\\
\addlinespace
Feasibility & \textcolor{black}{6.3\%} & \textcolor{black}{0.93} & \textcolor{black}{0.50} & \cellcolor[HTML]{F5FFFA}{0.48} & \textcolor{black}{0.93} & \cellcolor[HTML]{FFFFFF}{0.00} & \cellcolor[HTML]{7BB89C}{0.96} & \textcolor{black}{0.46} & \textcolor{black}{0.93} & \textcolor{black}{0.00} & \textcolor{black}{0.93} & \textcolor{black}{0.50} & \textcolor{black}{0.93} & \textcolor{black}{0.00} & \textcolor{black}{1.00}\\
\addlinespace
Suggestion & \textcolor{black}{4.5\%} & \textcolor{black}{0.97} & \textcolor{black}{0.55} & \cellcolor[HTML]{DDF1E8}{0.59} & \textcolor{black}{0.97} & \cellcolor[HTML]{FBFFFD}{0.19} & \cellcolor[HTML]{75B497}{0.98} & \textcolor{black}{0.82} & \textcolor{black}{0.97} & \textcolor{black}{0.67} & \textcolor{black}{0.97} & \textcolor{black}{0.55} & \textcolor{black}{0.97} & \textcolor{black}{0.11} & \textcolor{black}{1.00}\\
\addlinespace
Applicant: Quantity & \textcolor{black}{1.6\%} & \textcolor{black}{0.98} & \textcolor{black}{0.50} & \cellcolor[HTML]{F5FFFA}{0.50} & \textcolor{black}{0.98} & \cellcolor[HTML]{FFFFFF}{0.00} & \cellcolor[HTML]{73B395}{0.99} & \textcolor{black}{0.49} & \textcolor{black}{0.98} & \textcolor{black}{0.00} & \textcolor{black}{0.98} & \textcolor{black}{0.50} & \textcolor{black}{0.98} & \textcolor{black}{0.00} & \textcolor{black}{1.00}\\
\bottomrule
\end{tabular}
\end{table}

%% file: tables/table_f1_compare_detailed_multitask.tex
\begin{table}
\centering
\caption{\label{tab:tab:f1_compare_detailed_multitask}Detailed overview of the classification test results based on multi-task classification. Note:  Categories are sorted in descending order of label shares.}
\centering
\fontsize{6}{8}\selectfont
\begin{tabular}[t]{l>{\centering\arraybackslash}p{1cm}>{\centering\arraybackslash}p{1cm}>{\centering\arraybackslash}p{1cm}>{\centering\arraybackslash}p{1cm}>{\centering\arraybackslash}p{1cm}>{\centering\arraybackslash}p{1cm}>{\centering\arraybackslash}p{1cm}>{\centering\arraybackslash}p{1cm}>{\centering\arraybackslash}p{1cm}>{\centering\arraybackslash}p{1cm}>{\centering\arraybackslash}p{1cm}>{\centering\arraybackslash}p{1cm}>{\centering\arraybackslash}p{1cm}>{\centering\arraybackslash}p{1cm}>{\centering\arraybackslash}p{1cm}}
\toprule
\textbf{Category} & \textbf{Share Label} & \textbf{Acc.} & \textbf{Bal. Acc.} & \textbf{F1 (Macro)} & \textbf{F1 (Micro)} & \textbf{F1 Lab=1} & \textbf{F1 Lab=0} & \textbf{Prec. (Macro)} & \textbf{Prec. (Micro)} & \textbf{Prec. Lab=1} & \textbf{Prec. Lab=0} & \textbf{Recall (Macro)} & \textbf{Recall (Micro)} & \textbf{Recall Lab=1} & \textbf{Recall Lab=0}\\
\midrule
Proposal & \textcolor{black}{40.9\%} & \textcolor{black}{0.75} & \textcolor{black}{0.74} & \cellcolor[HTML]{B5D9C8}{0.74} & \textcolor{black}{0.75} & \cellcolor[HTML]{C3E2D3}{0.69} & \cellcolor[HTML]{A6D1BD}{0.80} & \textcolor{black}{0.74} & \textcolor{black}{0.75} & \textcolor{black}{0.71} & \textcolor{black}{0.77} & \textcolor{black}{0.74} & \textcolor{black}{0.75} & \textcolor{black}{0.66} & \textcolor{black}{0.82}\\
\addlinespace
Positive & \textcolor{black}{37.1\%} & \textcolor{black}{0.83} & \textcolor{black}{0.81} & \cellcolor[HTML]{A2CEBA}{0.81} & \textcolor{black}{0.83} & \cellcolor[HTML]{B1D7C5}{0.76} & \cellcolor[HTML]{94C6AF}{0.87} & \textcolor{black}{0.82} & \textcolor{black}{0.83} & \textcolor{black}{0.79} & \textcolor{black}{0.85} & \textcolor{black}{0.81} & \textcolor{black}{0.83} & \textcolor{black}{0.72} & \textcolor{black}{0.89}\\
\addlinespace
Applicant & \textcolor{black}{19.7\%} & \textcolor{black}{0.94} & \textcolor{black}{0.89} & \cellcolor[HTML]{8BC1A8}{0.90} & \textcolor{black}{0.94} & \cellcolor[HTML]{9CCBB5}{0.84} & \cellcolor[HTML]{7BB89C}{0.96} & \textcolor{black}{0.90} & \textcolor{black}{0.94} & \textcolor{black}{0.85} & \textcolor{black}{0.96} & \textcolor{black}{0.89} & \textcolor{black}{0.94} & \textcolor{black}{0.82} & \textcolor{black}{0.97}\\
\addlinespace
Rationale & \textcolor{black}{18.4\%} & \textcolor{black}{0.82} & \textcolor{black}{0.50} & \cellcolor[HTML]{F5FFFA}{0.45} & \textcolor{black}{0.82} & \cellcolor[HTML]{FFFFFF}{0.00} & \cellcolor[HTML]{8BC1A8}{0.90} & \textcolor{black}{0.41} & \textcolor{black}{0.82} & \textcolor{black}{0.00} & \textcolor{black}{0.82} & \textcolor{black}{0.50} & \textcolor{black}{0.82} & \textcolor{black}{0.00} & \textcolor{black}{1.00}\\
\addlinespace
Track Record & \textcolor{black}{18.0\%} & \textcolor{black}{0.95} & \textcolor{black}{0.91} & \cellcolor[HTML]{88BFA6}{0.91} & \textcolor{black}{0.95} & \cellcolor[HTML]{98C8B2}{0.85} & \cellcolor[HTML]{79B69A}{0.97} & \textcolor{black}{0.91} & \textcolor{black}{0.95} & \textcolor{black}{0.85} & \textcolor{black}{0.97} & \textcolor{black}{0.91} & \textcolor{black}{0.95} & \textcolor{black}{0.84} & \textcolor{black}{0.97}\\
\addlinespace
Relevance, Originality, Topicality & \textcolor{black}{17.1\%} & \textcolor{black}{0.86} & \textcolor{black}{0.60} & \cellcolor[HTML]{D1EADE}{0.64} & \textcolor{black}{0.86} & \cellcolor[HTML]{F8FFFB}{0.35} & \cellcolor[HTML]{84BDA3}{0.92} & \textcolor{black}{0.90} & \textcolor{black}{0.86} & \textcolor{black}{0.95} & \textcolor{black}{0.86} & \textcolor{black}{0.60} & \textcolor{black}{0.86} & \textcolor{black}{0.21} & \textcolor{black}{1.00}\\
\addlinespace
Methods & \textcolor{black}{16.2\%} & \textcolor{black}{0.84} & \textcolor{black}{0.51} & \cellcolor[HTML]{F5FFFA}{0.48} & \textcolor{black}{0.84} & \cellcolor[HTML]{FEFFFE}{0.05} & \cellcolor[HTML]{87BFA5}{0.91} & \textcolor{black}{0.92} & \textcolor{black}{0.84} & \textcolor{black}{1.00} & \textcolor{black}{0.84} & \textcolor{black}{0.51} & \textcolor{black}{0.84} & \textcolor{black}{0.02} & \textcolor{black}{1.00}\\
\addlinespace
Negative & \textcolor{black}{15.2\%} & \textcolor{black}{0.87} & \textcolor{black}{0.60} & \cellcolor[HTML]{D3EBE0}{0.63} & \textcolor{black}{0.87} & \cellcolor[HTML]{F8FFFB}{0.32} & \cellcolor[HTML]{83BCA2}{0.93} & \textcolor{black}{0.88} & \textcolor{black}{0.87} & \textcolor{black}{0.88} & \textcolor{black}{0.87} & \textcolor{black}{0.60} & \textcolor{black}{0.87} & \textcolor{black}{0.20} & \textcolor{black}{1.00}\\
\addlinespace
Suitability & \textcolor{black}{8.5\%} & \textcolor{black}{0.91} & \textcolor{black}{0.50} & \cellcolor[HTML]{F5FFFA}{0.48} & \textcolor{black}{0.91} & \cellcolor[HTML]{FFFFFF}{0.00} & \cellcolor[HTML]{7CB89C}{0.96} & \textcolor{black}{0.46} & \textcolor{black}{0.91} & \textcolor{black}{0.00} & \textcolor{black}{0.91} & \textcolor{black}{0.50} & \textcolor{black}{0.91} & \textcolor{black}{0.00} & \textcolor{black}{1.00}\\
\addlinespace
Feasibility & \textcolor{black}{6.3\%} & \textcolor{black}{0.94} & \textcolor{black}{0.50} & \cellcolor[HTML]{F5FFFA}{0.48} & \textcolor{black}{0.94} & \cellcolor[HTML]{FFFFFF}{0.00} & \cellcolor[HTML]{79B69A}{0.97} & \textcolor{black}{0.47} & \textcolor{black}{0.94} & \textcolor{black}{0.00} & \textcolor{black}{0.94} & \textcolor{black}{0.50} & \textcolor{black}{0.94} & \textcolor{black}{0.00} & \textcolor{black}{1.00}\\
\addlinespace
Suggestion & \textcolor{black}{4.5\%} & \textcolor{black}{0.96} & \textcolor{black}{0.50} & \cellcolor[HTML]{F5FFFA}{0.49} & \textcolor{black}{0.96} & \cellcolor[HTML]{FFFFFF}{0.00} & \cellcolor[HTML]{77B598}{0.98} & \textcolor{black}{0.48} & \textcolor{black}{0.96} & \textcolor{black}{0.00} & \textcolor{black}{0.96} & \textcolor{black}{0.50} & \textcolor{black}{0.96} & \textcolor{black}{0.00} & \textcolor{black}{1.00}\\
\addlinespace
Applicant: Quantity & \textcolor{black}{1.6\%} & \textcolor{black}{0.98} & \textcolor{black}{0.50} & \cellcolor[HTML]{F5FFFA}{0.50} & \textcolor{black}{0.98} & \cellcolor[HTML]{FFFFFF}{0.00} & \cellcolor[HTML]{73B395}{0.99} & \textcolor{black}{0.49} & \textcolor{black}{0.98} & \textcolor{black}{0.00} & \textcolor{black}{0.98} & \textcolor{black}{0.50} & \textcolor{black}{0.98} & \textcolor{black}{0.00} & \textcolor{black}{1.00}\\
\bottomrule
\end{tabular}
\end{table}

%% file: tables/table_f1_compare_detailed_binary_sentences_with_full_agreement_only.tex
\begin{table}
\centering
\caption{\label{tab:tab:f1_compare_detailed_binary (sentences with full agreement only)}Detailed overview of the classification test results based on binary (sentences with full agreement only) classification. Note:  Categories are sorted in descending order of label shares.}
\centering
\fontsize{6}{8}\selectfont
\begin{tabular}[t]{l>{\centering\arraybackslash}p{1cm}>{\centering\arraybackslash}p{1cm}>{\centering\arraybackslash}p{1cm}>{\centering\arraybackslash}p{1cm}>{\centering\arraybackslash}p{1cm}>{\centering\arraybackslash}p{1cm}>{\centering\arraybackslash}p{1cm}>{\centering\arraybackslash}p{1cm}>{\centering\arraybackslash}p{1cm}>{\centering\arraybackslash}p{1cm}>{\centering\arraybackslash}p{1cm}>{\centering\arraybackslash}p{1cm}>{\centering\arraybackslash}p{1cm}>{\centering\arraybackslash}p{1cm}>{\centering\arraybackslash}p{1cm}}
\toprule
\textbf{Category} & \textbf{Share Label} & \textbf{Acc.} & \textbf{Bal. Acc.} & \textbf{F1 (Macro)} & \textbf{F1 (Micro)} & \textbf{F1 Lab=1} & \textbf{F1 Lab=0} & \textbf{Prec. (Macro)} & \textbf{Prec. (Micro)} & \textbf{Prec. Lab=1} & \textbf{Prec. Lab=0} & \textbf{Recall (Macro)} & \textbf{Recall (Micro)} & \textbf{Recall Lab=1} & \textbf{Recall Lab=0}\\
\midrule
Proposal & \textcolor{black}{40.9\%} & \textcolor{black}{0.85} & \textcolor{black}{0.83} & \cellcolor[HTML]{9ACAB3}{0.84} & \textcolor{black}{0.85} & \cellcolor[HTML]{A4D0BC}{0.80} & \cellcolor[HTML]{91C4AC}{0.88} & \textcolor{black}{0.85} & \textcolor{black}{0.85} & \textcolor{black}{0.87} & \textcolor{black}{0.84} & \textcolor{black}{0.83} & \textcolor{black}{0.85} & \textcolor{black}{0.75} & \textcolor{black}{0.92}\\
\addlinespace
Positive & \textcolor{black}{37.1\%} & \textcolor{black}{0.92} & \textcolor{black}{0.91} & \cellcolor[HTML]{88BFA6}{0.91} & \textcolor{black}{0.92} & \cellcolor[HTML]{8DC2AA}{0.89} & \cellcolor[HTML]{82BCA1}{0.93} & \textcolor{black}{0.91} & \textcolor{black}{0.92} & \textcolor{black}{0.88} & \textcolor{black}{0.94} & \textcolor{black}{0.91} & \textcolor{black}{0.92} & \textcolor{black}{0.90} & \textcolor{black}{0.92}\\
\addlinespace
Applicant & \textcolor{black}{19.7\%} & \textcolor{black}{0.95} & \textcolor{black}{0.93} & \cellcolor[HTML]{85BEA4}{0.92} & \textcolor{black}{0.95} & \cellcolor[HTML]{92C5AD}{0.87} & \cellcolor[HTML]{79B69A}{0.97} & \textcolor{black}{0.91} & \textcolor{black}{0.95} & \textcolor{black}{0.84} & \textcolor{black}{0.98} & \textcolor{black}{0.93} & \textcolor{black}{0.95} & \textcolor{black}{0.91} & \textcolor{black}{0.96}\\
\addlinespace
Rationale & \textcolor{black}{18.4\%} & \textcolor{black}{0.85} & \textcolor{black}{0.67} & \cellcolor[HTML]{C0E0D1}{0.70} & \textcolor{black}{0.85} & \cellcolor[HTML]{F5FFFA}{0.49} & \cellcolor[HTML]{88BFA6}{0.91} & \textcolor{black}{0.75} & \textcolor{black}{0.85} & \textcolor{black}{0.63} & \textcolor{black}{0.88} & \textcolor{black}{0.67} & \textcolor{black}{0.85} & \textcolor{black}{0.40} & \textcolor{black}{0.95}\\
\addlinespace
Track Record & \textcolor{black}{18.0\%} & \textcolor{black}{0.95} & \textcolor{black}{0.90} & \cellcolor[HTML]{88BFA6}{0.91} & \textcolor{black}{0.95} & \cellcolor[HTML]{98C8B2}{0.85} & \cellcolor[HTML]{79B69A}{0.97} & \textcolor{black}{0.92} & \textcolor{black}{0.95} & \textcolor{black}{0.88} & \textcolor{black}{0.96} & \textcolor{black}{0.90} & \textcolor{black}{0.95} & \textcolor{black}{0.82} & \textcolor{black}{0.98}\\
\addlinespace
Relevance, Originality, Topicality & \textcolor{black}{17.1\%} & \textcolor{black}{0.92} & \textcolor{black}{0.85} & \cellcolor[HTML]{97C8B1}{0.85} & \textcolor{black}{0.92} & \cellcolor[HTML]{B1D7C5}{0.76} & \cellcolor[HTML]{7EB99E}{0.95} & \textcolor{black}{0.86} & \textcolor{black}{0.92} & \textcolor{black}{0.76} & \textcolor{black}{0.95} & \textcolor{black}{0.85} & \textcolor{black}{0.92} & \textcolor{black}{0.75} & \textcolor{black}{0.95}\\
\addlinespace
Methods & \textcolor{black}{16.2\%} & \textcolor{black}{0.90} & \textcolor{black}{0.78} & \cellcolor[HTML]{A5D0BC}{0.80} & \textcolor{black}{0.90} & \cellcolor[HTML]{CBE6D9}{0.66} & \cellcolor[HTML]{80BBA0}{0.94} & \textcolor{black}{0.83} & \textcolor{black}{0.90} & \textcolor{black}{0.74} & \textcolor{black}{0.92} & \textcolor{black}{0.78} & \textcolor{black}{0.90} & \textcolor{black}{0.59} & \textcolor{black}{0.96}\\
\addlinespace
Negative & \textcolor{black}{15.2\%} & \textcolor{black}{0.92} & \textcolor{black}{0.82} & \cellcolor[HTML]{9CCBB5}{0.83} & \textcolor{black}{0.92} & \cellcolor[HTML]{BBDDCD}{0.72} & \cellcolor[HTML]{7DB99D}{0.95} & \textcolor{black}{0.85} & \textcolor{black}{0.92} & \textcolor{black}{0.75} & \textcolor{black}{0.94} & \textcolor{black}{0.82} & \textcolor{black}{0.92} & \textcolor{black}{0.68} & \textcolor{black}{0.96}\\
\addlinespace
Suitability & \textcolor{black}{8.5\%} & \textcolor{black}{0.94} & \textcolor{black}{0.75} & \cellcolor[HTML]{A7D2BE}{0.79} & \textcolor{black}{0.94} & \cellcolor[HTML]{D7EDE3}{0.61} & \cellcolor[HTML]{79B69A}{0.97} & \textcolor{black}{0.86} & \textcolor{black}{0.94} & \textcolor{black}{0.76} & \textcolor{black}{0.96} & \textcolor{black}{0.75} & \textcolor{black}{0.94} & \textcolor{black}{0.51} & \textcolor{black}{0.98}\\
\addlinespace
Feasibility & \textcolor{black}{6.3\%} & \textcolor{black}{0.97} & \textcolor{black}{0.77} & \cellcolor[HTML]{9DCBB6}{0.83} & \textcolor{black}{0.97} & \cellcolor[HTML]{C5E3D5}{0.68} & \cellcolor[HTML]{75B497}{0.98} & \textcolor{black}{0.93} & \textcolor{black}{0.97} & \textcolor{black}{0.89} & \textcolor{black}{0.97} & \textcolor{black}{0.77} & \textcolor{black}{0.97} & \textcolor{black}{0.55} & \textcolor{black}{1.00}\\
\addlinespace
Suggestion & \textcolor{black}{4.5\%} & \textcolor{black}{0.98} & \textcolor{black}{0.79} & \cellcolor[HTML]{99C9B3}{0.85} & \textcolor{black}{0.98} & \cellcolor[HTML]{BFDFD0}{0.70} & \cellcolor[HTML]{74B396}{0.99} & \textcolor{black}{0.92} & \textcolor{black}{0.98} & \textcolor{black}{0.87} & \textcolor{black}{0.98} & \textcolor{black}{0.79} & \textcolor{black}{0.98} & \textcolor{black}{0.59} & \textcolor{black}{1.00}\\
\addlinespace
Applicant: Quantity & \textcolor{black}{1.6\%} & \textcolor{black}{0.99} & \textcolor{black}{0.87} & \cellcolor[HTML]{92C5AD}{0.87} & \textcolor{black}{0.99} & \cellcolor[HTML]{B3D8C7}{0.75} & \cellcolor[HTML]{72B294}{1.00} & \textcolor{black}{0.87} & \textcolor{black}{0.99} & \textcolor{black}{0.75} & \textcolor{black}{1.00} & \textcolor{black}{0.87} & \textcolor{black}{0.99} & \textcolor{black}{0.75} & \textcolor{black}{1.00}\\
\bottomrule
\end{tabular}
\end{table}

%% file: tables/table_f1_compare_rationale.tex
\begin{table}[h!]
\centering
\caption{\label{tab:tab:f1_compare_rationale}Comparison of F1 scores (macro-average) for detection of Rationale. The first row reports performance metrics for a model that uses only one sentence to detect Rationale. The second row  reports performance metrics for a model that adds the two surrounding sentences and combines the human annotation of Rationale and Rationale (Context).}
\centering
\fontsize{8}{10}\selectfont
\begin{tabular}[t]{>{\raggedright\arraybackslash}p{3cm}>{\centering\arraybackslash}p{1cm}>{\centering\arraybackslash}p{1cm}>{\centering\arraybackslash}p{1cm}>{\centering\arraybackslash}p{1cm}>{\centering\arraybackslash}p{1cm}>{\centering\arraybackslash}p{1cm}>{\centering\arraybackslash}p{1cm}>{\centering\arraybackslash}p{1cm}>{\centering\arraybackslash}p{1cm}}
\toprule
\textbf{Model} & \textbf{F1 (Macro)} & \textbf{F1 Lab=1} & \textbf{F1 Lab=0} & \textbf{Prec. (Macro)} & \textbf{Prec. Lab=1} & \textbf{Prec. Lab=0} & \textbf{Recall (Macro)} & \textbf{Recall Lab=1} & \textbf{Recall Lab=0}\\
\midrule
\textcolor{black}{Rationale (One Sentence Only)} & \textcolor{black}{0.71} & \textcolor{black}{0.52} & \textcolor{black}{0.90} & \textcolor{black}{0.71} & \textcolor{black}{0.53} & \textcolor{black}{0.89} & \textcolor{black}{0.7} & \textcolor{black}{0.50} & \textcolor{black}{0.90}\\
\addlinespace
\textcolor{black}{Rationale and Context} & \textcolor{black}{0.71} & \textcolor{black}{0.59} & \textcolor{black}{0.83} & \textcolor{black}{0.71} & \textcolor{black}{0.61} & \textcolor{black}{0.81} & \textcolor{black}{0.7} & \textcolor{black}{0.56} & \textcolor{black}{0.84}\\
\bottomrule
\end{tabular}
\end{table}

%% file: tables/table_f1_compare_binary_full.tex
\begin{table}[h!]
\centering
\caption{\label{tab:tab:f1_compare_binary_full}Comparison of F1-scores (macro-average) for binary classifiers and binary classifiers that only consider sentences with full agreement in the training set. Note: Categories are sorted in
descending order of label shares. Colours highlight the order and value of F1-scores. The last row shows the average F1 scores across all categories.}
\centering
\fontsize{9}{11}\selectfont
\begin{tabular}[t]{l>{\centering\arraybackslash}p{1.5cm}>{\centering\arraybackslash}p{3.5cm}>{\centering\arraybackslash}p{3.5cm}}
\toprule
\textbf{Category} & \textbf{Share Label} & \textbf{F1 Binary} & \textbf{F1 Binary \newline (Full Agreement)}\\
\midrule
Proposal & \textcolor{black}{40.9\%} & \cellcolor[HTML]{9ECCB7}{0.83} & \cellcolor[HTML]{9ACAB3}{0.84}\\
\addlinespace
Positive & \textcolor{black}{37.1\%} & \cellcolor[HTML]{8CC2A9}{0.89} & \cellcolor[HTML]{88BFA6}{0.91}\\
\addlinespace
Applicant & \textcolor{black}{19.7\%} & \cellcolor[HTML]{88BFA6}{0.91} & \cellcolor[HTML]{85BEA4}{0.92}\\
\addlinespace
Rationale & \textcolor{black}{18.4\%} & \cellcolor[HTML]{BEDFD0}{0.71} & \cellcolor[HTML]{C0E0D1}{0.70}\\
\addlinespace
Track Record & \textcolor{black}{18.0\%} & \cellcolor[HTML]{88BFA6}{0.91} & \cellcolor[HTML]{88BFA6}{0.91}\\
\addlinespace
Relevance, Originality, Topicality & \textcolor{black}{17.1\%} & \cellcolor[HTML]{97C8B1}{0.86} & \cellcolor[HTML]{97C8B1}{0.85}\\
\addlinespace
Methods & \textcolor{black}{16.2\%} & \cellcolor[HTML]{9FCDB8}{0.83} & \cellcolor[HTML]{A5D0BC}{0.80}\\
\addlinespace
Negative & \textcolor{black}{15.2\%} & \cellcolor[HTML]{9ECCB7}{0.83} & \cellcolor[HTML]{9CCBB5}{0.83}\\
\addlinespace
Suitability & \textcolor{black}{8.5\%} & \cellcolor[HTML]{A7D2BE}{0.79} & \cellcolor[HTML]{A7D2BE}{0.79}\\
\addlinespace
Feasibility & \textcolor{black}{6.3\%} & \cellcolor[HTML]{93C5AE}{0.87} & \cellcolor[HTML]{9DCBB6}{0.83}\\
\addlinespace
Suggestion & \textcolor{black}{4.5\%} & \cellcolor[HTML]{A0CDB8}{0.82} & \cellcolor[HTML]{99C9B3}{0.85}\\
\addlinespace
Applicant: Quantity & \textcolor{black}{1.6\%} & \cellcolor[HTML]{83BCA2}{0.93} & \cellcolor[HTML]{92C5AD}{0.87}\\
\addlinespace
\midrule
Average F1 Score Across all Categories & \textcolor{black}{---} & \cellcolor[HTML]{99C9B3}{0.85} & \cellcolor[HTML]{9ACAB3}{0.84}\\
\bottomrule
\end{tabular}
\end{table}

%% file: tables/table_f1_compare_detailed_llama.tex
\begin{table}
\centering
\caption{\label{tab:llama_annotation}Detailed overview of the classification test results based on few-shot learning using a Large Language Model (Meta-Llama-3-8B-Instruct). Note:  Categories are sorted in descending order of label shares.}
\centering
\fontsize{6}{8}\selectfont
\begin{tabular}[t]{l>{\centering\arraybackslash}p{1cm}>{\centering\arraybackslash}p{1cm}>{\centering\arraybackslash}p{1cm}>{\centering\arraybackslash}p{1cm}>{\centering\arraybackslash}p{1cm}>{\centering\arraybackslash}p{1cm}>{\centering\arraybackslash}p{1cm}>{\centering\arraybackslash}p{1cm}>{\centering\arraybackslash}p{1cm}>{\centering\arraybackslash}p{1cm}>{\centering\arraybackslash}p{1cm}>{\centering\arraybackslash}p{1cm}>{\centering\arraybackslash}p{1cm}>{\centering\arraybackslash}p{1cm}>{\centering\arraybackslash}p{1cm}}
\toprule
\textbf{Category} & \textbf{Share Label} & \textbf{Acc.} & \textbf{Bal. Acc.} & \textbf{F1 (Macro)} & \textbf{F1 (Micro)} & \textbf{F1 Lab=1} & \textbf{F1 Lab=0} & \textbf{Prec. (Macro)} & \textbf{Prec. (Micro)} & \textbf{Prec. Lab=1} & \textbf{Prec. Lab=0} & \textbf{Recall (Macro)} & \textbf{Recall (Micro)} & \textbf{Recall Lab=1} & \textbf{Recall Lab=0}\\
\midrule
Proposal & \textcolor{black}{40.9\%} & \textcolor{black}{0.68} & \textcolor{black}{0.65} & \cellcolor[HTML]{CBE6D9}{0.66} & \textcolor{black}{0.68} & \cellcolor[HTML]{E0F3EA}{0.58} & \cellcolor[HTML]{B6DAC9}{0.74} & \textcolor{black}{0.66} & \textcolor{black}{0.68} & \textcolor{black}{0.62} & \textcolor{black}{0.71} & \textcolor{black}{0.65} & \textcolor{black}{0.68} & \textcolor{black}{0.54} & \textcolor{black}{0.77}\\
\addlinespace
Positive & \textcolor{black}{37.1\%} & \textcolor{black}{0.79} & \textcolor{black}{0.83} & \cellcolor[HTML]{A9D3C0}{0.79} & \textcolor{black}{0.79} & \cellcolor[HTML]{ADD5C2}{0.77} & \cellcolor[HTML]{A5D0BC}{0.80} & \textcolor{black}{0.81} & \textcolor{black}{0.79} & \textcolor{black}{0.64} & \textcolor{black}{0.98} & \textcolor{black}{0.83} & \textcolor{black}{0.79} & \textcolor{black}{0.98} & \textcolor{black}{0.67}\\
\addlinespace
Applicant & \textcolor{black}{19.7\%} & \textcolor{black}{0.91} & \textcolor{black}{0.90} & \cellcolor[HTML]{93C5AE}{0.87} & \textcolor{black}{0.91} & \cellcolor[HTML]{A5D0BC}{0.80} & \cellcolor[HTML]{7FBA9F}{0.94} & \textcolor{black}{0.85} & \textcolor{black}{0.91} & \textcolor{black}{0.73} & \textcolor{black}{0.97} & \textcolor{black}{0.90} & \textcolor{black}{0.91} & \textcolor{black}{0.88} & \textcolor{black}{0.92}\\
\addlinespace
Rationale & \textcolor{black}{18.4\%} & \textcolor{black}{0.79} & \textcolor{black}{0.53} & \cellcolor[HTML]{EDFAF3}{0.53} & \textcolor{black}{0.79} & \cellcolor[HTML]{FBFFFD}{0.18} & \cellcolor[HTML]{90C4AC}{0.88} & \textcolor{black}{0.58} & \textcolor{black}{0.79} & \textcolor{black}{0.33} & \textcolor{black}{0.83} & \textcolor{black}{0.53} & \textcolor{black}{0.79} & \textcolor{black}{0.12} & \textcolor{black}{0.95}\\
\addlinespace
Track Record & \textcolor{black}{18.0\%} & \textcolor{black}{0.92} & \textcolor{black}{0.89} & \cellcolor[HTML]{93C5AE}{0.87} & \textcolor{black}{0.92} & \cellcolor[HTML]{A8D2BF}{0.79} & \cellcolor[HTML]{7EB99E}{0.95} & \textcolor{black}{0.86} & \textcolor{black}{0.92} & \textcolor{black}{0.75} & \textcolor{black}{0.96} & \textcolor{black}{0.89} & \textcolor{black}{0.92} & \textcolor{black}{0.83} & \textcolor{black}{0.94}\\
\addlinespace
Relevance, Originality, Topicality & \textcolor{black}{17.1\%} & \textcolor{black}{0.79} & \textcolor{black}{0.75} & \cellcolor[HTML]{C0E0D1}{0.70} & \textcolor{black}{0.79} & \cellcolor[HTML]{ECF9F3}{0.53} & \cellcolor[HTML]{94C6AF}{0.87} & \textcolor{black}{0.68} & \textcolor{black}{0.79} & \textcolor{black}{0.43} & \textcolor{black}{0.93} & \textcolor{black}{0.75} & \textcolor{black}{0.79} & \textcolor{black}{0.69} & \textcolor{black}{0.81}\\
\addlinespace
Methods & \textcolor{black}{16.2\%} & \textcolor{black}{0.85} & \textcolor{black}{0.66} & \cellcolor[HTML]{C5E3D5}{0.68} & \textcolor{black}{0.85} & \cellcolor[HTML]{F6FFFA}{0.44} & \cellcolor[HTML]{87BFA5}{0.91} & \textcolor{black}{0.72} & \textcolor{black}{0.85} & \textcolor{black}{0.55} & \textcolor{black}{0.89} & \textcolor{black}{0.66} & \textcolor{black}{0.85} & \textcolor{black}{0.37} & \textcolor{black}{0.94}\\
\addlinespace
Negative & \textcolor{black}{15.2\%} & \textcolor{black}{0.78} & \textcolor{black}{0.86} & \cellcolor[HTML]{BCDECE}{0.71} & \textcolor{black}{0.78} & \cellcolor[HTML]{E1F3EB}{0.57} & \cellcolor[HTML]{98C8B2}{0.85} & \textcolor{black}{0.70} & \textcolor{black}{0.78} & \textcolor{black}{0.41} & \textcolor{black}{0.99} & \textcolor{black}{0.86} & \textcolor{black}{0.78} & \textcolor{black}{0.97} & \textcolor{black}{0.75}\\
\addlinespace
Suitability & \textcolor{black}{8.5\%} & \textcolor{black}{0.92} & \textcolor{black}{0.55} & \cellcolor[HTML]{E0F3EA}{0.58} & \textcolor{black}{0.92} & \cellcolor[HTML]{FBFFFD}{0.20} & \cellcolor[HTML]{7CB89C}{0.96} & \textcolor{black}{0.77} & \textcolor{black}{0.92} & \textcolor{black}{0.62} & \textcolor{black}{0.92} & \textcolor{black}{0.55} & \textcolor{black}{0.92} & \textcolor{black}{0.12} & \textcolor{black}{0.99}\\
\addlinespace
Feasibility & \textcolor{black}{6.3\%} & \textcolor{black}{0.94} & \textcolor{black}{0.68} & \cellcolor[HTML]{BEDFD0}{0.71} & \textcolor{black}{0.94} & \cellcolor[HTML]{F6FFFA}{0.44} & \cellcolor[HTML]{79B69A}{0.97} & \textcolor{black}{0.74} & \textcolor{black}{0.94} & \textcolor{black}{0.52} & \textcolor{black}{0.96} & \textcolor{black}{0.68} & \textcolor{black}{0.94} & \textcolor{black}{0.39} & \textcolor{black}{0.98}\\
\addlinespace
Suggestion & \textcolor{black}{4.5\%} & \textcolor{black}{0.94} & \textcolor{black}{0.77} & \cellcolor[HTML]{BADCCC}{0.72} & \textcolor{black}{0.94} & \cellcolor[HTML]{F5FFFA}{0.47} & \cellcolor[HTML]{79B69A}{0.97} & \textcolor{black}{0.69} & \textcolor{black}{0.94} & \textcolor{black}{0.39} & \textcolor{black}{0.98} & \textcolor{black}{0.77} & \textcolor{black}{0.94} & \textcolor{black}{0.59} & \textcolor{black}{0.96}\\
\addlinespace
Applicant: Quantity & \textcolor{black}{1.6\%} & \textcolor{black}{0.87} & \textcolor{black}{0.87} & \cellcolor[HTML]{E7F6EF}{0.55} & \textcolor{black}{0.87} & \cellcolor[HTML]{FBFFFD}{0.17} & \cellcolor[HTML]{83BCA2}{0.93} & \textcolor{black}{0.55} & \textcolor{black}{0.87} & \textcolor{black}{0.10} & \textcolor{black}{1.00} & \textcolor{black}{0.87} & \textcolor{black}{0.87} & \textcolor{black}{0.88} & \textcolor{black}{0.87}\\
\bottomrule
\end{tabular}
\end{table}